\documentstyle[aps,prl]{revtex}
\input{epsf.sty}
\input{psfig.sty}

\begin{document}


\title{The physics of dynamical atomic charges~: \\
           the case of ABO$_3$ compounds}
\author{Ph. Ghosez~\cite{Add}, J.-P. Michenaud and X. Gonze}
\address{Unit\'e de Physico-Chimie et de Physique des Mat\'eriaux,
Universit\'e Catholique de Louvain,\\
1 Place Croix du Sud, B-1348 Louvain-la-Neuve, Belgium}
\date{\today}
\maketitle
\begin{abstract}
Based on recent first-principles computations in perovskite compounds, especially BaTiO$_3$, we
examine the significance of the Born effective charge concept and contrast it with other atomic
charge definitions, either static (Mulliken, Bader...) or dynamical (Callen, Szigeti...). It is shown
that static and dynamical charges are not driven by the same underlying parameters. A unified
treatment of dynamical charges in periodic solids and large clusters is proposed. 
The origin of the difference between static and dynamical
charges is discussed in terms of local polarizability and delocalized transfers of charge~: 
local models succeed in reproducing anomalous effective charges thanks to 
large atomic polarizabilities but, in ABO$_3$ compounds, ab initio
calculations favor the physical picture based upon transfer of charges. Various results concerning
barium and strontium titanates are presented. The origin of anomalous Born effective charges is
discussed thanks to a band-by-band decomposition
which allows to identify the displacement of the Wannier center of separated bands induced by an
atomic displacement. The sensitivity of the Born effective charges to microscopic and
macroscopic strains is examined. Finally, we estimate the spontaneous polarization in the four
phases of barium titanate.
\end{abstract}

\setcounter{page}{1}

\section{Introduction}
\label{Intro}

For a long time, there has been a continuing interest in the definition of atomic charges in
solid state physics as well as in chemistry~\cite{Cochran61,Mulliken35,Wiberg93,Meister94}.
This interest lies essentially in the fact that the concept of atomic charge naturally arises
in a large diversity of frameworks and is frequently helpful for a simple description of solids
and molecules. The variety of contexts in which the charge is involved (IR spectrum analysis, XPS
chemical shifts analysis, theory of ionic conductivity of oxides, determination of electrostatic
potential, definition of oxidation states...) underlines its central role but also reveals a
concomitant problem: inspired by various models or by the  description of various physical
phenomena, many different  definitions have been proposed that, unfortunately, are not
equivalent~\cite{Meister94}. 

Following a distinction already made by Cochran~\cite{Cochran61}, it seems possible to
classify the different concepts into static and dynamical charges. The {\it static} charge is an
intuitive concept, usually based on a partitioning of the ground-state electronic density into
contributions attributed to the different atoms. It is an ill-defined quantity that depends on the
convention artificially chosen to affect a  given electron to a particular
ion~\cite{Cochran61,Mulliken35}. On the other hand, the {\it dynamical} charge is directly
related to the change of polarization  (or dipole moment, for molecules) created by an atomic
displacement. This change of polarization is a quantity that can be experimentally measured,
at least in principles, giving the dynamical charge a well-defined character.

In order to clarify the concept of atomic charge, it was important to compare on practical
examples the numerical results obtained from its different definitions. Recent studies of the
statistical correlation between various atomic charges using a principal component analysis have
suggested that the different definitions are not independent but correspond to different scales
driven by a unique underlying physical factor~\cite{Meister94}. If this assertion seems plausible
as far as static charges are concerned, we will argue that the dynamical charge should not reduce
to the same physical factor but should also depend on an additional parameter: the rate of
transfer of charge, influenced by the bonding with the other atoms of the system and additionally,
for large systems, by the condition imposed on the macroscopic electric field.

The Born effective charge tensor $Z^{*(T)}$ -- alias transverse charge --, that is at the center of the
present study,  is a dynamical quantity introduced by  Born~\cite{Born33} in 1933.  In solid
state physics, it is since a long time considered as  a fundamental quantity because it
monitors the long-range Coulomb interaction responsible of the splitting between transverse
and longitudinal optic phonon modes~\cite{Born33}. During the seventies, the Born effective
charges were already investigated and discussed within empirical approaches (see for
example Harrison~\cite{Harrison80}). More recently, they became accessible to first-principles
calculations~\cite{Baroni87,Gonze92,KingSmith93}, and accurate values have been reported
for a large variety of materials. 

For the case of ABO$_3$ compounds, old experimental data~\cite{Axe67} and empirical
studies~\cite{Harrison80} had suggested that the amplitude of the Born effective charges should
deviate substantially from the amplitude of the static atomic charge. Surprisingly, this result
remained in the dark until first-principles calculations confirmed that the components of
$Z^{*(T)}$ are anomalously large in these oxides~\cite{Resta93,Ghosez94,Zhong94}. It was observed
that the components of $Z^{*(T)}$ can reach twice that of the nominal ionic charges. This result
reopened the discussion on the physics of the Born effective charges and different recent studies
tried to clarify the microscopic processes monitoring the amplitude of $Z^{*(T)}$. 

In this paper, we first clarify the relationship between various atomic charges. We then 
present results concerning BaTiO$_3$ and SrTiO$_3$ in
order to illustrate how a careful analysis of the Born effective charges can teach us interesting
physics concerning these compounds. It reveals the mixed ionic and covalent character of the
bond~\cite{Ghosez95a,Posternak94}. It allows to visualize the mechanism of polarization as
electronic currents produced by dynamical changes of orbital
hybridizations~\cite{Harrison80,Posternak94}. It also clarifies the origin of the giant
destabilizing dipole-dipole interaction producing the ferroelectric instability of these
materials~\cite{Ghosez96}.

In Section II and III, we contrast the concepts of static
and dynamical charges and we reintroduce the Born effective charge that is at the
center of the present discussion.
In Section IV, we compare various results obtained within different
frameworks for the cubic phase of BaTiO$_3$ and SrTiO$_3$. The bond orbital model 
of Harrison is explicitely applied to SrTiO$_3$ (Appendix A). We also discuss
the origin of the large anomalous contributions in terms of local electronic 
polarizability and dynamical changes of orbital hybridization. A decomposition of 
the role played by the different bands is reported in Section V.
Section VI is devoted to the evolution of the Born effective charges in
the three ferroelectric phases of  BaTiO$_3$ as well as in the cubic phase
under hydrostatic pressure.  This points out the role of the anisotropy of the
atomic environment on the amplitude of $Z^{*(T)}$. Finally, in Section VII, we report
the evolution of the effective charges all along the path of  atomic 
displacements from the cubic to the rhombohedral phase and we estimate the
spontaneous polarization of the three ferroelectric phase of BaTiO$_3$.

\section {The concept of static charge}
\label{Static}

Intuitively, the atomic charge may first appear as a static concept.
The charge associated to an isolated atom is a well-defined quantity. The purpose of
defining static atomic charges is therefore to extend this notion to molecules
and solids. For these cases, the challenge basically consists to replace
the delocalized electronic density by localized point charges associated to
each atom.  This could {\it a priori} be performed from electronic density
maps obtained experimentally or theoretically. However, as  already
mentioned by Mulliken~\cite{Mulliken35} in 1935, ``there are some
difficulties of giving exact definition without arbitrariness for any atomic
property''. During the seventies, Cochran~\cite{Cochran61} similarly
emphasized that the partition of the electronic distribution into atomic
charges can only be done  unambiguously when ``boundary can be drawn
between the ions so as to pass through regions in which the electron density
is small compared with the reciprocal of the  volume inclosed''. This is
never the case in practice, and especially when there is appreciable covalent
bonding. For most of the solids and molecules, there is consequently no
{\it absolute} criterion to define the static atomic charge and a large variety of
distinct definitions have been proposed that are not necessarily quantatively equivalent (see for
instance Ref.~\cite{Wiberg93,Meister94}).

As an illustration, different approaches have been considered in order to evaluate the
amplitude of the static atomic charges of barium titanate. Some results are summarized in
Table~\ref{Zcubic},  where different atomic charges are reported in comparison with the nominal
charges expected in a purely ionic material  (+2 for Ba, +4 for Ti, --2 for O).  Some of them were
obtained from empirical models; others were deduced from first-principles. The static atomic
charges of Ref.~\cite{Harrison80} were deduced by Harrison within his bond orbital model using
universal parameters and neglecting the interactions with the Ba atom. The atomic charges
reported by Hewat~\cite{Hewat73} were approximated from a model initially designed by
Cowley~\cite{Cowley64} for SrTiO$_3$. The charges reported by Khatib {\it et
al.}~\cite{Khatib89} have been obtained in a shell-model context. In two references, Turik and
Khasabov~\cite{Turik88,Turik95} estimated the charges from the Madelung constant thanks to a
fit of the crystal energy with shell-model parameters. Michel-Calendini {\it et
al.}~\cite{Michel80} proposed charges from a population analysis of the X$\alpha$ electronic
distribution of a TiO$_6$ cluster, assuming a charge of $+2$ on Ba. Cohen and
Krakauer~\cite{Cohen90} deduced the atomic charges from a fit of the DFT-LDA  electronic
distribution  by that of overlapping spherical ions (generated according to the potential induced
breathing model) for different ionic configurations.  Xu {\it et al.}~\cite{Xu90} reported values
deduced from a Mulliken population analysis of a self-consistent OLCAO
calculation~\cite{Note_M}. In another reference~\cite{Xu94}, Xu {\it et al.} proposed different
values by integrating the electronic charges in spheres centered on the ions, and partitioning
rather arbitrarily  the remaining charge outside the spheres following a method proposed  in
Ref.~\cite{Ching90,Xu91}.

The results of Table~\ref{Zcubic} are not quantitatively identical and illustrate that there is no
formal equivalence between the different procedures used to define the atomic charge. However, in
agreement with an analysis reported by Meister and Schwartz~\cite{Meister94} for the case of
molecules, we observe that the values of Table~\ref{Zcubic} have some common features,
suggesting that the different charges are not independent but should correspond to different
scales driven by a common factor.

\begin{figure}[tb]
\caption{Ti-O plane cut in the partial electronic density coming from the O 2p bands 
in the cubic phase of barium titanate.}
\label{Fig.1}
\vspace{2cm}
\centerline{\psfig{figure={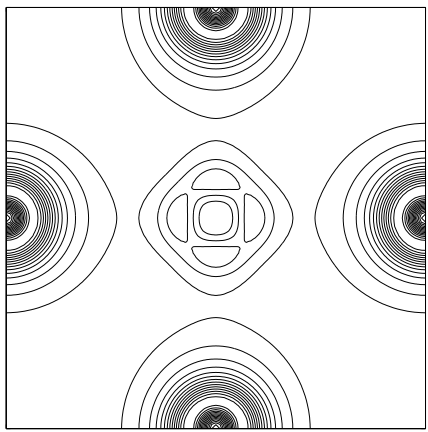},angle=0,height=6.0cm}}
\vspace{4cm}
\end{figure}

In particular, all the calculations reveal that the charge transfer from Ti to O is not complete. If
BaTiO$_3$ was a purely ionic crystal, the 3d and 4s electrons of Ti would be entirely transferred
to the oxygen atoms, yielding a charge of +4 on titanium. However, due for instance to the partial 
hybridization between O 2p and Ti 3d
states~\cite{Nemoshkalenko85,Hudson93,Mattheiss72,Pertosa78a,Pertosa78b,Weyrich85a,Cohen92a},
these electrons remain partly delocalized on the Ti atom so that the static charges on the
Ti and O atoms are smaller than they would be in a purely ionic material. This delocalization 
is illustrated in Figure \ref{Fig.1}, where we have plotted the partial electronic density 
associated to the O 2p bands. For the  Ba atom, the situation is not so clear than for titanium
but most of the studies suggest similarly that the 6s electrons are not fully transferred to 
the oxygen.

{}From the previous examples, it is clear that, strictly speaking, the static charge does not give a
quantitative information. In the study of mixed ionic-covalent compounds, it remains however a
useful concept to discuss qualitatively the transfer of charges from one atom to the other. As a
general rule, the partial covalence reduces the amplitude of the static charge. Comparison of a
specific static charge in the different phases of a given material or in different compounds can
therefore give a relevant information on the evolution of the chemical bond~\cite{Xu94}.

\section {The concept of dynamical charge}
\label{Dynamical}

As emphasized by Harrison~\cite{Harrison80}, ``whenever an ambiguity arises about the definition
of a concept such as the atomic  charge, it can be removed by discussing only quantities that can be
experimentally  determined, at least in principles''. If there are some ambiguities to determine
the charge to be associated to a given atom, the charge carried by this atom when it is displaced is
directly accessible from the induced change of polarization (or dipole moment for molecules).
As it is now discussed, the dynamical charges are defined by the change of polarization induced by
an atomic displacement; from the viewpoint of Harrison, they appear therefore as  more
fondamental quantities. 

     \subsection{Role of the macroscopic electric field}
\label{Field}

In molecules, the change of dipole moment in direction $\beta$ ($p_{\beta}$) linearly induced by 
a small displacement of atom $\kappa$ in direction $\alpha$ ($\tau_{\kappa, \alpha}$) is uniquely 
defined. The proportionality coefficient
between the dipole moment and the atomic displacement has the dimentionality of a charge and is
usually referred to as the {\it atomic polar tensor} (APT)~:
\begin{eqnarray}
Z^*_{\kappa, \alpha \beta} &=& 
 \frac{\partial  {p}_{\beta}}{\partial \tau_{\kappa, \alpha}}
\label{Zmol}
\end{eqnarray}
This concept was introduced by Biarge, Herranz and
Morcillo~\cite{Biarge61,Morcillo66,Morcillo69} for the interpretation of infra-red intensities
measurements.  Later, Cioslowski~\cite{Cioslowski89a,Cioslowski89b} introduced a scalar
charge  in connection with this tensor~: it is the generalized atomic polar tensor (GAPT) defined
as one-third of the trace of the atomic polar tensor.

In periodic systems, equivalent atoms appear in the different unit cells and the definition of
the charge can be generalized. A dynamical charge tensor is conventionally defined as the 
coefficient of proportionality at the linear order between the {\it
macroscopic polarization} per unit cell created in direction $\beta$ and  a rigid displacement of
the {\it sublattice} of atoms $\kappa$ in direction $\alpha$, times the unit cell volume $\Omega_0$~: 
\begin{eqnarray}
Z^*_{\kappa, \alpha \beta} &=& 
 \Omega_0 \; \frac{\partial  {\cal P}_{\beta}}{\partial \tau_{\kappa, \alpha}}
\end{eqnarray}
We note that $\Omega_0 . {\cal P}$ can be interpreted as a dipole moment per unit cell. As
one $\kappa$ atom is displaced in each unit cell, in the linear regime, this definition is
equivalent to Eq. (\ref{Zmol})~: it corresponds to the change of dipole moment induced by an
isolated atomic displacement. However, contrary to the case of molecules, in macroscopic
systems, the previous quantity is not uniquely defined. Indeed, the change of polarization
is also dependent on the boundary conditions fixing the macroscopic electric field ${\cal E}$
throughout the sample. Basically, we can write~:
\begin{eqnarray}
Z^*_{\kappa, \alpha \beta} &=& 
\Omega_0 \; {\left. \frac{\partial  {\cal P}_{\beta}}{\partial \tau_{\kappa, \alpha}}
\right|}_{{\cal E}=0}
+ 
\Omega_0 \; \sum_{j} 
\frac{\partial  {\cal P}_{\beta}}{\partial {\cal E}_{j}} 
\; . \; 
\frac{\partial  {\cal E}_{j}}{\partial \tau_{\kappa, \alpha}}
\end{eqnarray}
As the electrostatics imposes a relationship between macroscopic polarization, electric
and displacement fields~:
\begin{eqnarray}
\label{A4}
{\cal D}_{\alpha} =  {\cal E}_{\alpha}+ 4 \pi {\cal P}_{\alpha}
                           =  \sum_{j} {\epsilon}^{\infty}_{\alpha, j} {\cal E}_j
\end{eqnarray}
we can deduce the following equivalent expression~: 
\begin{eqnarray}
\label{A5}
Z^*_{\kappa, \alpha \beta} &=& 
\Omega_0 \; {\left. \frac{\partial  {\cal P}_{\beta}}{\partial \tau_{\kappa, \alpha}}
\right|}_{{\cal E}=0}
+ 
\Omega_0 \;
 \sum_{j} 
\frac{({\epsilon}^{\infty} _{\beta, j} - \delta_{\beta, j})}
                                       {4 \pi}
\; . \;  
\frac{\partial  {\cal E}_{j}}{\partial \tau_{\kappa, \alpha}}
\end{eqnarray}
Depending on the condition imposed on the macroscopic electric field, different concepts have
historically been introduced~\cite{Note_Gervais}.

The {\it Born effective charge}~\cite{Born33} -- alias transverse charge, $Z^{*(T)}$
-- refers to the change of polarization that would be observed under the condition of zero
macroscopic electric field, so that the second term appearing in the previous equation vanishes~:
\begin{equation}
Z^{*(T)}_{\kappa, \alpha \beta} = \Omega_0 \; 
{\left. \frac{\partial  {\cal P}_{\beta}}{\partial \tau_{\kappa, \alpha}}
\right|}_{{\cal E}=0}
\end{equation}

The {\it Callen charge}~\cite{Callen49} -- alias longitudinal charge, $Z^{*(L)}$ -- is
defined from the change of polarization under the condition of zero macroscopic displacement
field ~:
\begin{equation}
Z^{*(L)}_{\kappa, \alpha \beta} = \Omega_0 \; 
{\left. \frac{\partial  {\cal P}_{\beta}}{\partial \tau_{\kappa, \alpha}}
\right|}_{{\cal D}=0} .
\end{equation}

Introducing in Eq. (\ref{A5}) the relationship between field ${\cal E}$ and 
polarization ${\cal P}$, deduced from Eq.
(\ref{A4}) under the condition of vanishing displacement field, Born and Callen charges can be
related to each other thanks to the knowledge of the optical dielectric tensor
$\epsilon^{\infty}$~:
\begin{eqnarray}
Z^{*(L)}_{\kappa, \alpha \beta}
&=&
Z^{*(T)}_{\kappa, \alpha \beta}
- 
 \sum_{j} 
\frac{({\epsilon}^{\infty} _{\beta, j} - \delta_{\beta, j})}
                                       {4 \pi}
\; . \;  4 \pi
\underbrace{ 
\Omega_0 \;
{\left. \frac{\partial  {\cal P}_{j}}{\partial \tau_{\kappa, \alpha}}
\right|}_{{\cal D}=0} 
}_{Z^{*(L)}_{\kappa, \alpha j}}
\end{eqnarray}
so that finally~\cite{Martin81}~:
\begin{eqnarray}
Z^{*(T)}_{\kappa, \alpha \beta}
&=&
 \sum_{j} {\epsilon}^{\infty} _{\beta j} \; \; Z^{*(L)}_{\kappa, \alpha j}
\end{eqnarray}
For the case of isotropic materials, we recover the well known equality~: $Z^{*(T)}_{\kappa}=
\epsilon^{\infty} . Z^{*(L)}_{\kappa}$. Even if they are both related to the change of polarization
induced by an atomic displacement, Born and Callen charges appear as two distinct
quantities and  will be significantly different in materials where
$\epsilon^{\infty}$ is different from unity. 

Basically, an infinite number of charges could be defined corresponding to different boundary
conditions, relating ${\cal P}$ and ${\cal E}$. One of them is the {\it Szigeti
charge}~\cite{Szigeti49,Burstein67} -- $Z^{*(S)}$ --, defined as the change of polarization under the
condition of vanishing local field, ${\cal E}_{loc}$, at the atomic site:
\begin{equation}
Z^{*(S)}_{\kappa, \alpha \beta} =
\Omega_0 \;  {\left. \frac{\partial  {\cal P}_{\beta}}{\partial \tau_{\kappa, \alpha}}
\right|}_{{\cal E}_{loc}=0}  
\end{equation}
Contrary to Born and Callen charges, $Z^{*(S)}$ was sometimes considered as a model-dependent
concept in the sense that the local field is not observable as the macroscopic field but require
some assumptions to be estimated.  In the particular case of an isotropic material,  the condition
of vanishing local field can be written as follows: 
\begin{eqnarray}
{\cal E}_{loc} = {\cal E} + \; \frac{4 \pi}{3} {\cal P} = 0
\end{eqnarray}
Introducing this condition in Eq. (\ref{A5})~:
\begin{eqnarray}
Z^{*(S)}_{\kappa}
&=&
Z^{*(T)}_{\kappa}
- 
\frac{({\epsilon}^{\infty} - 1)}
                                       {4 \pi}
\; . \;  \frac{4 \pi}{3}
\underbrace{ 
\Omega_0 \; 
{\left. \frac{\partial  {\cal P}_{}}{\partial \tau_{\kappa}}
\right|}_{{\cal E}_{loc}= 0} 
}_{Z^{*(S)}_{\kappa}}
\end{eqnarray}
so that we find~:
\begin{eqnarray}
\label{Eq4}
Z^{*(T)}_{\kappa}
&=&
 \frac{({\epsilon}^{\infty}+2)}{3} \; \; Z^{*(S)}_{\kappa}
\end{eqnarray}

In calculations of the dynamical properties of crystals, the contribution from the long-range
Coulombic interaction to the atomic forces is usually restricted to dipolar forces and is included
through a term~:  $F^d_{\kappa}= Z^{*(T)}_{\kappa}  {\cal E}$. {}From Eq.~(\ref{Eq4}), it can be checked
that this force can be alternatively written in terms of local quantities~: 
$F^d_{\kappa}= Z^{*(S)}_{\kappa} {\cal E}_{loc}$. In shell-model calculations, this second formulation 
is usually preferred. Indeed, from its definition, $Z^{*(S)}$ only includes the effects of charge
redistribution resulting from short-range interactions and it is therefore conveniently assimilated
to the static charge~\cite{Cochran60,Michel80}.  

{}From the previous discussion, it appears that the amplitude of the dynamical charge in
macroscopic bodies is sensitive to the condition imposed on the macroscopic electric field.
Considering finite clusters of increasing size, we deduce that the amplitude of the dynamical
charge, reducing to the APT for a microscopic body, will tend to  a different value when the
macroscopic limit is taken, depending from the {\it shape} of the cluster.  We investigate now this
observation in more details, and provide a unified treatment
of dynamical charges in periodic solids and clusters, sufficiently large
for the macroscopic quantities (${\cal E}, {\cal P}, \epsilon^{\infty}, ...$) to be defined.

Following the well-known practice for the study of dielectric bodies~\cite{Landau69},
we consider that the cluster has a macroscopic ellipso\"{\i}dal shape. 
In this case, the macroscopic field within the cluster present
the practical advantage to be homogeneous. In absence of any applied external field, it reduces to
the depolarizing field related to the macroscopic polarization thanks to the depolarization
coefficients $n_{\alpha}$~\cite{Landau69}~. If we assume in what follows that the principal 
axes of the ellipso\"{\i}d are aligned with the axes of coordinates, we have the following
relationship~:
\begin{eqnarray}
{\cal E}_{\alpha} = - {4 \pi} n_{\alpha} {\cal P}_{\alpha} 
\end{eqnarray}
where the geometry imposes~: $\sum_i n_{i}=1$. Following the same procedure as previously, the
dynamical charge $Z^{*(E)}$ of a given atom $\kappa$ in an ellipso\"{\i}d of volume $\Omega$ can be
written as~:
\begin{eqnarray}
Z^{*(E)}_{\kappa, \alpha \beta}
&=&
\Omega \; {\left. \frac{\partial  {\cal P}_{\beta}}{\partial \tau_{\kappa, \alpha}}
\right|}_{{\cal E}=0}
+ 
\Omega \;
 \sum_{j} 
\frac{({\epsilon}^{\infty} _{\beta, j} - \delta_{\beta, j})}
                                       {4 \pi}
\; . \;  
{\left.  \frac{\partial  {\cal E}_{j}}{\partial \tau_{\kappa, \alpha}} 
\right|}_{{\cal E}_j=- 4 \pi n_j {\cal P}_j}  \\
&=&
Z^{*(T)}_{\kappa, \alpha \beta}
- 
 \sum_{j} 
\frac{({\epsilon}^{\infty} _{\beta, j} - \delta_{\beta, j})}
                                       {4 \pi}
\; . \;  4 \pi n_j
\underbrace{ 
\Omega \;
{\left. \frac{\partial  {\cal P}_{j}}{\partial \tau_{\kappa, \alpha}}
\right|}_{{\cal E}_j=- 4 \pi n_j {\cal P}_j} 
}_{Z^{*(E)}_{\kappa, \alpha j}}
\end{eqnarray}
and we have the general relationship~:
\begin{eqnarray}
Z^{*(T)}_{\kappa, \alpha \beta}
&=&
 \sum_{j} [ ({\epsilon}^{\infty} _{\beta j} - \delta_{\beta j}) n_j 
                   +   \delta_{\beta j} ] \; \; Z^{*(E)}_{\kappa, \alpha j}
\end{eqnarray}
In this expression, the presence of the depolarization coefficients emphasizes the influence of
the shape of the cluster on the amplitude of $Z^{*(E)}$. 
The above-mentioned sum rule on the depolarization coefficients forbid to impose the
condition of zero electric or displacement fields simultaneously in the three directions.
However, we have the following three interesting cases. 

First, we consider an extremely oblate ellipso\"{\i}dal (slab-like) cluster 
and take the macroscopic limit. Along the $z$ direction 
perpendicular to the surface, $n_{z} \rightarrow 1$, while, along the two other directions,
$n_{x}= n_{y} \rightarrow 0$.
The dynamical charge for the ellipso\"{\i}d is therefore related to the Born effective 
charge through the following expression:
\begin{eqnarray}
\left(
   \begin{array}{c}
    Z^{*(T)}_{\kappa, \alpha x} \\
    Z^{*(T)}_{\kappa, \alpha y} \\
    Z^{*(T)}_{\kappa, \alpha z} \\
   \end{array}
\right) 
= 
\left(
   \begin{array}{ccc}
    1   & 0  & \epsilon^{\infty}_{xz}  \\
    0   & 1  & \epsilon^{\infty}_{yz}  \\
    0   & 0  & \epsilon^{\infty}_{zz}  \\
   \end{array}
\right) 
\left(
   \begin{array}{c}
    Z^{*(E)}_{\kappa, \alpha x} \\
    Z^{*(E)}_{\kappa, \alpha y} \\
    Z^{*(E)}_{\kappa, \alpha z} \\
   \end{array}
\right) 
\end{eqnarray}
For uniaxial systems with no off-diagonal terms in the dielectric tensor, we note that the cluster 
charge along the direction perpendicular to the slab becomes identified with the Callen charge,
while that in the slab plane reduces to the Born effective charge.

Differently, for an extremely prolate ellipso\"{\i}dal (needle-like) cluster 
aligned along the $z$ direction (for which $n_{z}
\rightarrow 0$ and $n_{x}= n_{y} \rightarrow 1/2$), we have the following relationship:
\begin{eqnarray}
\left(
   \begin{array}{c}
    Z^{*(T)}_{\kappa, \alpha x} \\
    Z^{*(T)}_{\kappa, \alpha y} \\
    Z^{*(T)}_{\kappa, \alpha z} \\
   \end{array}
\right) 
= 
\left(
   \begin{array}{ccc}
    \frac{1}{2} (\epsilon^{\infty}_{xx}+1)  & \frac{1}{2} \epsilon^{\infty}_{xy} & 0 \\
    \frac{1}{2} \epsilon^{\infty}_{yx}  & \frac{1}{2} (\epsilon^{\infty}_{yy}+1)   & 0 \\
    \frac{1}{2} \epsilon^{\infty}_{zx}  & \frac{1}{2} \epsilon^{\infty}_{zy}  & 1 \\
   \end{array}
\right) 
\left(
   \begin{array}{c}
    Z^{*(E)}_{\kappa, \alpha x} \\
    Z^{*(E)}_{\kappa, \alpha y} \\
    Z^{*(E)}_{\kappa, \alpha z} \\
   \end{array}
\right) 
\end{eqnarray}
Here also, the charge along the $z$ direction will reduce to the Born charge in
uniaxial systems.

Finally, for a spherical cluster, the symmetry imposes $n_1 = n_2 = n_3 = 1/3$, so that 
${\cal E}_{\alpha}= - 4 \pi {\cal P} / 3$. 
For the case of an isotropic material, we recover therefore the
condition of vanishing local field and $Z_{\kappa}^{(E)}$ becomes equivalent to $Z_{\kappa}^{(S)}$.
Therefore, we obtain the interesting result that in isotropic compounds, 
the Szigeti charge appears as a well-defined quantity and is 
simply the dynamical charge observed in a spherical cluster.

To summarize, the concept of dynamical charge in macroscopic systems is not uniquely defined~: it
depends on the relationship between ${\cal E}$ and ${\cal P}$. In each case, the charge was 
however expressed in terms of two basic concepts, $Z^{*(T)}$ and $\epsilon^{\infty}$. In this Section,
we focused on the term that includes the dielectric constant, and that describes the part of the 
electronic charge redistribution induced by the presence of a macroscopic field. In the next Section, 
we will discuss the physical processes responsible of the amplitude of $Z^{*(T)}$.

     \subsection{Dynamical changes of orbital hybridizations}
\label{Hybrid}

During the seventies, a large variety of semi-empirical models were proposed to investigate the
underlying physical processes driving the amplitude of dynamical charges. Without being exhaustive, let us
mention the interesting treatments of Lucovsky, Martin and Burnstein~\cite{Lucovsky71} who
decomposed $Z^{*(T)}$ in a local and a non-local contribution, of Lucovsky and 
White~\cite{Lucovsky73}
discussing $Z^{*(T)}$ in connection with resonant bonding properties, or the bond charge model of
H\"ubner~\cite{Hubner75}. The most popular and sophisticated of these approaches remains
however that of Harrison~\cite{Harrison80,Harrison73,Harrison74a,Harrison74b} within his bond
orbital model (BOM). A similar theory was developed independently by Lannoo
and Decarpigny~\cite{Lannoo73}. 

The BOM basically consists in a simplified tight-binding model, where the
Hamiltonian is limited to the on-site and nearest-neighbour terms. The on-site
elements are identified to free atom terms value, while the interatomic
elements are taken as universal constants times a particular distance
dependence. Among other things, these parameters determine the transfer of
charge between the interacting atoms. As noted  by
Dick and Overhauser~\cite{Dick58}, the charge redistribution produced by the
sensitivity of the overlap integrals on the atomic positions is at the origin
of an ``exchange charge polarization''. Similarly, in the Harrison model, the
dependence of the parameter on the bond length are at the origin of dynamical
transfer of charges and monitors the amplitude of $Z^{*(T)}$ that can become 
anomalously large as it is illustrated in the following examples.

Let us first consider a diatomic molecule XY, composed of two open
shell atoms, where Y has the largest electronegativity. The interatomic
distance is $u$ and the dipole moment $p(u)$. These observables allow us to identify
a convenient static charge $Z(u)=\frac{p(u)}{u}$, while the dynamical charge is defined as~:
\begin{eqnarray}
\label{Zmod}
Z^{*}(u) &= &\frac{\partial p(u)}{\partial u}   \nonumber \\
            &= &\frac{\partial }{\partial u} \; \big( u \: . \: Z(u) \big)  
\nonumber \\
            &= & Z(u)+u\frac{\partial Z(u)}{\partial u}
\end{eqnarray}
In the last expression, $Z^*$ appears composed of two terms. The first one is 
simply the static charge. The second corresponds to an additional {\it dynamical} 
contribution: it originates in the transfer of charge produced by the modification 
of the interatomic distance. Within the BOM, this last contribution is 
associated to off-site orbital hybridization changes and is deduced 
from the universal dependence of the orbital interaction parameters on the 
bond length, as illustrated on a practical example in Appendix~\ref{Apdx-2}. 
We deduce that the difference 
between $Z(u)$ and $Z^{*}(u)$ will be large if $Z(u)$ changes rapidly with $u$. 
It can even be non-negligible when  ${\partial p(u)}/{\partial u}$ is small, 
if the charge is transferred on a large distance $u$.

Moreover, this simple model naturally predicts {\it anomalous} amplitude of the dynamical
charges, i.e. a value of $Z^*(u)$ not only larger than the static charge $Z(u)$ but
even larger than the ``nominal'' ionic charge. As the distance between X and Y
is modified from 0 to some $\overline{u}$, the distance corresponding to a
{\it complete} transfer of electrons from X to Y,  the dipole moment evolves
continuously from
$p(0)=0$ (since there is no dipole for that case) to $p(\overline{u})$.  
Interestingly,
\begin{equation}
\int^{\overline{u}}_{0} Z^{*}(u) \; du = [p(\overline{u})-p(0)]
          = \overline{u} \; Z(\overline{u}) \\
\end{equation}
so that:
\begin{equation}
\frac{1}{\overline{u}}\int^{\overline{u}}_{0} Z^{*}(u) du = Z(\overline{u})
\end{equation}
{}From the last relationship
the mean value of $Z^{*}(u)$ from 0 to $\overline{u}$ is equal to
$Z(\overline{u})$ (the ``nominal'' ionic charge). Consequently, if $Z(u)$ changes with $u$,
$Z^{*}(u)$ must be larger than $Z(\overline{u})$ for some $u$ between
$[0,\overline{u}]$. The difference between $Z^{*}(u)$ and the
nominal charge $Z(\overline{u})$ is usually referred to as the {\it
anomalous} contribution~\cite{Note_1}.

Switching now from a molecule to a linear chain ...-Y-X-Y-..., and displacing
coherently the X atoms by $du$, shortened and elongated bonds will alternate all along the
chain. For Harrison~\cite{Harrison80}, the interaction parameters will be
modified such that ``the covalent energy increases in the shorted bond,
making it less polar by transferring electron to the positive atom''.
Inversely, electronic charge will be transferred to the negative atom in the
elongated bond. These transfers of charge will propagate all along the chain,
so that even if the net charge on the atom is not modified, a {\it current} of
electrons will be associated to the atomic displacement. {\it The direction of
this electronic current is opposite to that of the displacement of positive
atoms, so that it reinforces the change of polarization associated to this
displacement and may generate an anomalously large dynamical charge}. 
In our example, we have implicitely considered a truly periodic system under the 
condition of zero macroscopic electric field so that
the associated dynamical charge is $Z^{*(T)}$. Under other conditions, the amplitude 
of the transfers of charge would be additionally influenced by the presence of the 
field as discussed in the previous Section. We note that, contrary to what was observed 
for the static charge, consequences of the covalence effects
are to increase the amplitude of $Z^{*(T)}$.

The previous model can finally be extended to three dimensional solids. For
this case, however, the calculation of the dynamical contribution may become
questionable when the identification of the charge transfers is restricted to
some specific bonds~\cite{Bennetto96}.  As it will be discussed in Sections~\ref{Cubic}
and~\ref{BbB} the Harrison model remains however a meaningful picture of practical
interest to interpret more accurate results.

Up to now, we focused on a ``delocalized'' model within which the electronic charge
redistribution induced by an atomic displacement is visualized by transfer of charge
induced by {\it off-site} changes of hybridization. In the past, various shell-models
have however also been developed to investigate the dynamical properties of crystals. In 
these calculations, an accurate description of $Z^{*(T)}$ was mandatory in order to 
reproduce correctly the splitting between longitudinal and transverse optic modes in
the vicinity of the $\Gamma$ point. Contrary to the BOM, the shell-model is ``local'' and
treats the  charges within the Clausius-Mosotti limit. The previous dicussion in terms of a
static and dynamical contribution to $Z^{*(T)}$ remains valid. 
However, the dynamical contribution results there simply from the relative displacement 
of the shell charge as a whole with respect to the atom. It is attributed to the 
polarizability of the electrons in the local field at the atomic site. In the
language of the BOM, such a displacement of the electronic cloud can be understood in
terms of {\it on-site} changes of hybridizations. This approach contrasts with the model
developped by Harrison but can also yield plausible Born effective charge 
amplitudes~/cite{Ghosez95b}. 

\begin{figure}[tb]
\caption{Schematic representation of the two basic mechanisms that
can explain the displacement of the Wannier center of a band under 
atomic displacement~: (a) local polarizability, (b) interatomic
transfers of charge.}
\label{Fig.2}
\vspace{2cm}
\centerline{\psfig{figure={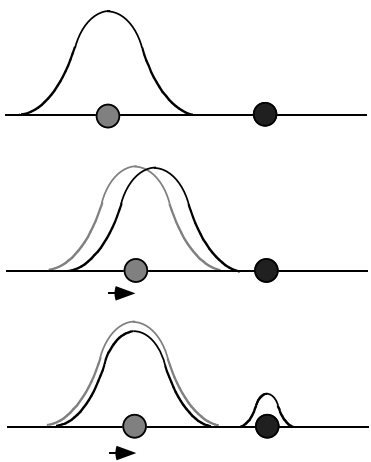},angle=0,height=6.0cm}}
\vspace{4cm}
\end{figure}

It must be emphasized that it is not possible to discriminate {\it a priori} between 
localized and delocalized models.
Within the recent theory of polarization, it has been clarified that for the purpose of
understanding polarization problems, ``the true quantum mechanical electronic system can be
considered as an effective classical system of quantized point charges, located at the centers of
gravity associated with the occupied Wannier functions in each unit cell''~\cite{Vanderbilt93}. 
Consequently,  the correct description of the Born effective charges does not require to reproduce
correctly  all the features of the valence charge distribution but {\it only} the displacement of its
Wannier center (see Ref.~\cite{Ghosez1}).  As schematized in Fig.~\ref{Fig.2}, antagonist models 
can reproduce a similar displacement of the Wannier center. In real materials, both local 
polarizability and transfers of charge do probably contribute to the charge reorganisation. 
It will be emphasized later how first-principles investigations can help to identify the 
dominant mechanism.

In conclusion, this Section has shown that $Z^*$ is related to the static charge (see Eq. \ref{Zmod})
but does not restrict to it: $Z^*$ may also include an additional, important, dynamical
contribution. Whatever the mechanism of the charge redistribution (localized or delocalized), the
amplitude of the dynamical contribution cannot be estimated from the inspection of the electronic
density alone.  So, we partly disagree  with Meister and Schwarz~\cite{Meister94} who suggested
that all the charges {\it including} the GAPT are driven by the same underlying parameter.   
In what follows, based on firs-principles calculations, 
we illustrate on different examples that $Z^{*(T)}$ may become anomalously large
and independent of the amplitude of the static charge $Z$. Moreover, two atoms with similar $Z$ 
can also exhibit strongly different $Z^{*(T)}$.

\section {A first-principles approach} 
\label{Technical}

A brief review of the most commonly used first-principles approaches 
for computing the Born effective charges has 
been reported recently~\cite{Ghosez1}. {}Going beyond semi-empirical approaches, 
{\it ab initio} techniques allow 
accurate prediction of $Z^{*(T)}$ in materials where its amplitude is not necessarily directly 
accessible from the experiment.
Going further, the first-principles approaches are also offering a new opportunity to clarify the
microscopic mechanism modulating the amplitude of $Z^{*(T)}$ without any preliminary hypothesis. As
it will be illustrated in the following sections, it reveals particularly  useful to understand 
the origin of anomalously large $Z^{*(T)}$ in ABO$_3$ compounds.

The results of the present paper have been obtained in the framework of the density
functional formalism~\cite{Pickett89,Payne92}. The exchange-correlation energy has been
evaluated within the local density approximation, using a Pad\'e
parametrization~\cite{TeterU} of Ceperley-Alder homogeneous electron gas
data~\cite{Ceperley80}. Integrals over the Brillouin-zone were replaced by a sum on a mesh of $6
\times 6 \times 6$ special {\bf k}-points~\cite{Monkhorst76,Monkhorst77} (10 points in the
irreducible Brillouin zone). The ``all electron'' potentials were replaced by the same ab initio,
separable, extended norm-conserving pseudopotentials as in Ref.~\cite{Ghosez94}. The
wavefunctions were expanded in plane waves up to a kinetic energy cutoff of 35 Hartree (about
4100 plane waves).

The Born effective charges have been deduced from linear response calculations~\cite{Baroni87}, using a
variational formulation~\cite{Gonze92,Gonze97a,Gonze97b} to the density functional perturbation
theory. The decomposition of individual contributions from separate groups of
occupied bands  has been performed following the scheme described in Ref.~\cite{Ghosez1}. The
parameters used for the calculations guarantee a convergency better than 0.5\% on $Z^{*(T)}$ as well
as on  each of its band-by-band contributions.

\section {The cubic phase of ABO$_3$ compounds}
\label{Cubic}

     \subsection{General results}

Recently, the Born effective charge tensors of perovskite ABO$_3$ compounds  have  been
at the center of numerous investigations
\cite{Resta93,Ghosez94,Zhong94,Ghosez95a,Posternak94,Ghosez95b,Rabe94,Yu95,Wang96b,LaSota97}.
In the cubic phase, they are fully characterized by a set of four independent
numbers. The charge tensor of the A and B atoms is isotropic owing to the local
spherical symmetry at the atomic site. For oxygen, the local environment is tetragonal and
two independent elements O$_{\parallel}$ and O$_{\perp}$ must be considered, referring
respectively to the change of polarization induced by an atomic displacement parallel and
perpendicular to the B-O bond. In Table~\ref{Z*cubicBa}, 
we summarize the results obtained within different approaches
for the cubic phase of BaTiO$_3$.  

The first reliable estimation of $Z^{*(T)}$ in BaTiO$_3$ is probably due to Axe~\cite{Axe67}, from
empirical fitting to experimental mode oscillator strengths~\cite{Note_Z}. In ABO$_3$
compounds, $Z^{*(T)}$ cannot be determined unambiguously from the experiment. However,
within some realistic hypothesis, Axe identified the independent elements of the effective
charges of BaTiO$_3$ and already pointed out their two essential features. First, the oxygen
charge tensor is highly anisotropic. Second, the charges on Ti and O$_{\parallel}$ contain a large
{\it anomalous} contribution (i.e. an additional charge with respect to the nominal ionic value of
+2 for Ba, +4 for Ti and -2 for O).

Both these characteristics are confirmed by the first-principles calculations.
Our ab initio results, computed from linear response, are also in excellent
agreement with those of Zhong {\it et al.}~\cite{Zhong94}, obtained from
finite differences of polarization. The charge neutrality sum rule,
reflecting the numerical accuracy of our calculation, is fulfilled 
to within 0.02. We note that the values of $Z^{*(T)}$ are also qualitatively reproduced from
a shell-model calculation~\cite{Ghosez95b} and accurately predicted within the SCAD
model~\cite{Boyer96}.  

The anomalous amplitude of the dynamical charge, reported in this Section, is not a specific
feature of BaTiO$_3$.  Similar computations of $Z^{*(T)}$ were performed on different perovskite
ABO$_3$ compounds and they all reproduce the same characteristics than in BaTiO$_3$.
A non exhaustive list of these results is reported in Table \ref{Z*cubicAB}.
We observe that the choice of the A
atom has a rather limited influence on $Z^{*(T)}_{B}$ and $Z^{*(T)}_{O \parallel}$,
which appear closely related to the B atom. While the nominal ionic charge of
Ti and Zr is +4 in these compounds, the Born effective charge is between
+$7.08$ and +$7.56$ for Ti, and approximately equal to +$6.03$ for Zr. For Nb,
the ionic charge is +5, while the Born effective charge is between +$9.11$
and +$9.37$. Extending the investigations to WO$_3$ in the reference cubic phase (defect
perovskite structure), the ionic charge on W is equal to +6, while the Born effective charge reaches
the much larger value of +$12.51$. For the class of perovskite ABO$_3$compounds, it can be checked
that $Z^{*(T)}_B$ evolves quasi linearly with the nominal charge of the B atom~\cite{Detraux97}.

For materials containing Pb, the previous considerations remain valid but
there are additional anomalies concerning $Z^{*(T)}_A$ and $Z^{*(T)}_{O \perp}$. This
feature is due to the more covalent bonding of lead with oxygen that was
illustrated in Ref.~\cite{Cohen92a,Cohen92b}. In what follows, we will not
be concerned with these lead compounds.

     \subsection{Origin of the anomalous contributions}

The approximate reciprocity between O$_{\parallel}$  and B anomalous
contributions suggests that they should originate in a global transfer of
charge between B and O atoms as described in Section~\ref{Hybrid}.  In
Ref.~\cite{Harrison80}, Harrison had in fact already suggested the existence of giant Born
effective charges in perovskite materials. Being unaware of the earlier
results of Axe, he had however no experimental evidence to corroborate his
semi-empirical calculations. 

In Appendix~\ref{Apdx-2}, we report results obtained within the Harrison model (it follows the
method described for KCl in Ref.~\cite{Harrison80}, p. 334.).  For SrTiO$_3$, from the
universal tight-binding parameters of Harrison, we get a value of $-$8.18 for $Z^{*(T)}_{O \parallel}$,
making plausible the giant anomalous effective charges only by focusing on the dynamical changes
of hybridization between occupied O 2{s}--O 2{p} states and the unoccupied metal { d} states. 
In BaTiO$_3$, the hybridization between O 2p and Ti 3d orbitals is a well known feature,
confirmed by various sources (experiments~\cite{Nemoshkalenko85,Hudson93}, LCAO calculations
\cite{Mattheiss72,Pertosa78a,Pertosa78b} and DFT results~\cite{Weyrich85a,Cohen92a}). In this
context, it was therefore realistic to focus on  O 2{p} - B {d} hybridization changes to
explain intuitively large anomalous contributions~\cite{Zhong94}. 

At the opposite, it may therefore appear surprizing that model calculations which does {\it not}
explicitly include transfer of charges are able to predict correctly the amplitude of the Born
effective charges. For instance, in Table I, we observe that the values of $Z^{*(T)}$ 
are qualitatively
reproduced by a shell-model calculation~\cite{Ghosez95b}. A similar agreement between {\it ab
initio} and shell model results was highlighted for KNbO$_3$~\cite{Spliarsky95}.  In both cases, 
the calculation was performed within the ``polarizability model'' introduced by Bilz
{\it et al.}~\cite{Bilz87}, which includes an anisotropic and non-linear polarizability of the O
atoms. In the same spirit, at the level of the SCAD model, the Born effective charges are accurately
reproduced while there is no explicit transfer of electrons between the different atomic sites.
As discussed in Section \ref{Dynamical}, antagonist models can be invoked to explain the origin of
anomalous contributions as soon as they globally reproduce a similar displacement of the
Wannier center of the valence charge distribution. What appear as a macroscopic current along the
Ti--O chain within the BOM shows itself as an unusual polarizability of the oxygen atoms within
the shell model.

It was not possible to discriminate unambiguously between localized and delocalized model
until Posternak {\it et al.}~\cite{Posternak94}  proposed a convincing proof of the 
crucial role of ``off-site'' hybridizations. Based on first-principles  calculations,
they demonstrated for KNbO$_3$ that the anomalous contribution to the charge of Nb and O$_{\parallel}$ 
disappears if the hybridization between O 2p and Nb 4d orbitals is artificially suppressed. 
In a similar spirit,  the inspection of the Wannier functions of BaTiO$_3$ and the analysis of
their deformation under an atomic displacement reported by Marzari and
Vanderbilt~\cite{MarzariU} confirm the predominant role played by the Ti 3d orbitals and the
explanation introduced by Harrison.

In the next Section, we propose a band-by-band decomposition of the Born effective
charges~\cite{Ghosez95a,Ghosez95b}. This technique appears as a tool of paramount
importance to clarify the microscopic origin of anomalous contributions. Identifying the
dynamical transfer of charges without any preliminary hypothesis on the orbitals that interact, 
it will allow to generalize the basic mechanism that was proposed by Harrison.

\section {Identification of dynamical changes of hybridization}
\label{BbB}

In ABO$_3$ compounds, the electronic band structure is composed of well separated sets of bands.
The hybridizations between the orbitals of the different atoms are relatively small and each
band can be identified by the name of the main atomic orbital which contributes to this energy
level in the solid. The Born effective charge is defined by the change of polarization associated to
a specific atomic displacement. Our purpose will be to identify the contribution of each
well separated set of bands to this change of polarization~\cite{Ghosez95a,Ghosez95b}.

     \subsection{Reference configuration}

In Ref.~\cite{Ghosez1}, we have described how band-by-band contributions to $Z^{*(T)}$ can be
separated from each others. Moreover, it has been demonstrated that the contribution 
to $Z^{*(T)}_{\kappa, \alpha \beta}$ from a
single occupied band $n$ can be interpreted as a change of polarization $\Omega_o \Delta{\cal
P}_{\beta}= -2 . \Omega_o \Delta d_{\beta}$ where $\Delta d_{\beta}$ is the displacement in
direction $\beta$ of the  Wannier center of band $n$, induced by the unitary displacement of the
sublattice of atoms $\kappa$ in  direction $\alpha$.                                  

In order to understand the origin of the displacement of the Wannier center of each band, it is
helpful to define a  {\it reference} configuration that corresponds to what we would expect in a
purely ionic material. In such fictitious material, each band would be composed of a single
non-hybridized orbital and the Wannier center of each band would be centered on a given atom. In
absence of any hybridization, when displacing a given sublattice of atoms $\kappa$, the Wannier
center of the bands centered on the moving atoms would remain centered on it, while the position
of the center of gravity of the other bands would remain unaffected. The contributions of these two
kind of bands to $Z^{*(T)}_{\kappa}$ would therefore be $-2$ and 0 electrons, respectively.

In the real material, the {\it anomalous} contribution of a particular band $m$ to a given atom
$\kappa$ is defined as the additional part with respect to the reference value
expected in absence of any hybridization: it reflects how the center of the Wannier center of
band $m$ is displaced relatively to the atoms when the sublattice $\kappa$ moves~\cite{Note_W}.  
Considering each band as a combinaison of atomic orbitals, such a  displacement of the Wannier center 
of a band with respect to its reference position {\it must} be attributed to hybridization effects: 
it is associated to  the admixture of a new orbital character to the band. When the orbitals which
interact are located on different atoms (``off-site'' hybridization), the dynamical changes of
hybridization can be visualized as transfers of charge. If the interacting orbitals are on the same
atom (``on-site'' hybridization), the mechanism much looks like a polarizability.

Rigorously, our band-by-band decomposition is performed within DFT and formally only concerns
the Kohn-Sham particles. It seems however that the results are rather independent of the
one-particle scheme~\cite{Massidda95} used for the calculation so that the results presented
here should give a good insight on the physics of the ABO$_3$ compounds.

     \subsection{BaTiO$_3$}

\begin{figure}[tb]
\caption{Kohn-Sham electronic band structure of BaTiO$_3$.}
\label{Fig.b.Ba}
\vspace{2cm}
\centerline{\psfig{figure={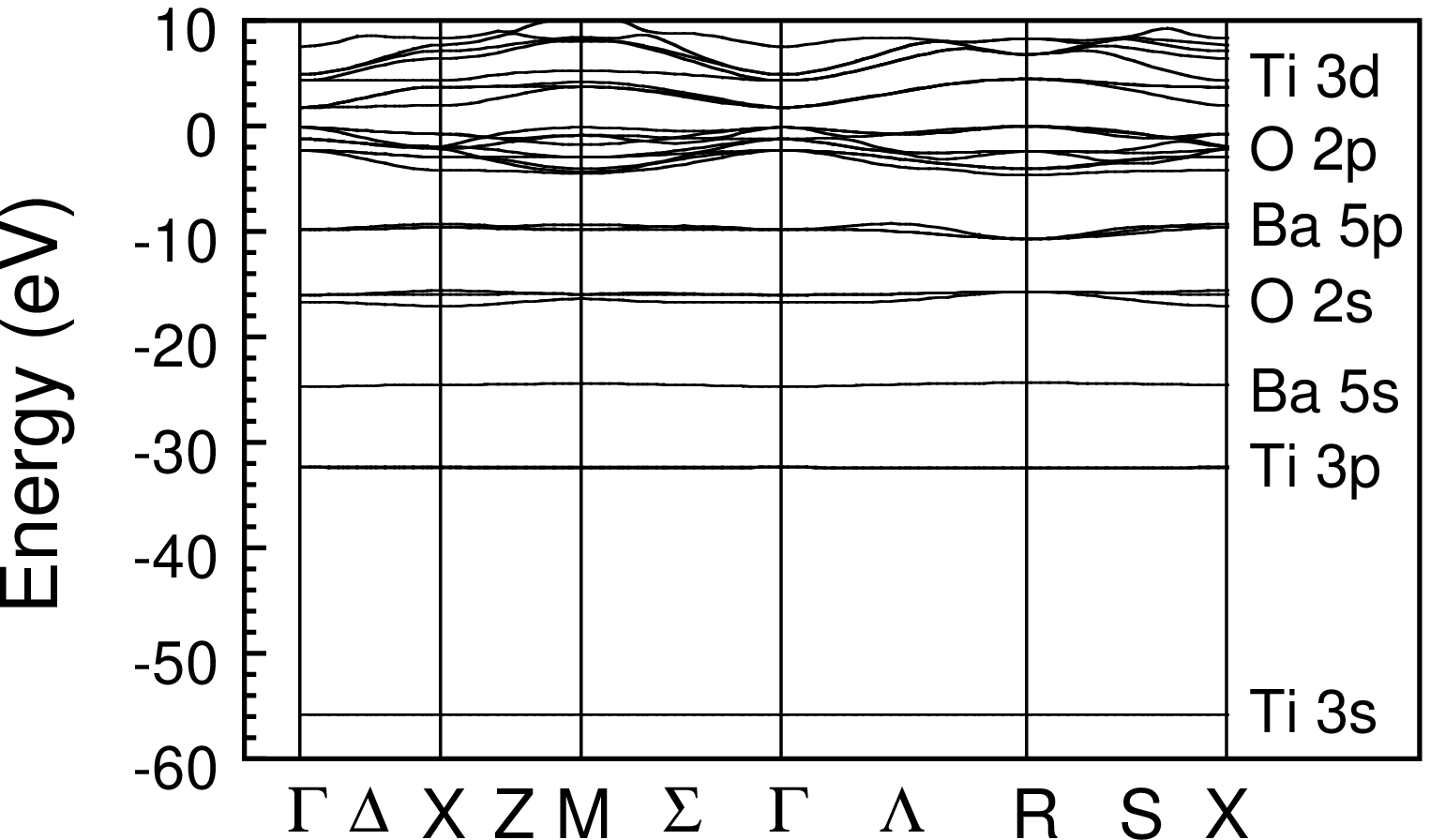},angle=0,height=6.0cm}}
\vspace{4cm}
\end{figure}

Let us first applied the band-by-band decomposition to barium titanate. The band structure of
BaTiO$_3$ is presented in Fig.~\ref{Fig.b.Ba}. Results of the decomposition of $Z^{*(T)}$
in the theoretical cubic structure of BaTiO$_3$ are reported in
Table~\ref{Z*bbbBa}. 
The first line ($Z_{\kappa}$) brings together the charge of the nucleus and
core electrons included in the pseudopotential. The other contributions
come from the different valence electron levels. The sum of the band-by-band
contributions on one atom is equal to its global effective charge while the
sum of the contribution to a particular band from the different atoms is equal
to $-2$ (within the accuracy of the calculation), the occupancy of this band.

Focusing first on the titanium charge, we observe that the Ti 3{s} contribution ($-2.03$) is close
to $-2$, confirming that the Ti 3{s} electrons follow the Ti atom when moving, independently
from the change of its surrounding. This result {\it a posteriori} justifies the inclusion of
deeper electronic levels as part of the ionic pseudopotentials. At the
opposite, it is shown that the giant anomalous charge of titanium essentially
comes from the O 2p bands ($+2.86$). It corresponds to a displacement of the Wannier
center of the O 2p bands in opposite direction to the displacement of the Ti
atom. This observation is in perfect agreement with
the Harrison model: it can be understood by dynamical changes of
hybridization between O 2p  and Ti 3d  orbitals, producing a transfer of
electron from O to Ti when the Ti-O distance shortens. This explanation was confirmed
recently from the inspection of the O 2p Wannier functions~\cite{MarzariU}. Beyond the previous
observations, we note however that there are also small anomalous charges from the Ti 3p, O 2s 
and Ba 5p  bands. These contributions are not negligible. The positive anomalous charges
correspond to a displacement of the center of the Wannier function of the O and Ba bands in the
direction of the closest Ti when this atom has moved. Some of these features go beyond the
Harrison  model, within which anomalous contributions to 
$Z^{*(T)}_{Ti}$ in Table ~\ref{Z*bbbBa} would be restricted to the O 2p and O 2s bands. They
suggest other kind of hybridization changes, that will be
now more explicitly investigated.

Focusing on barium, the global anomalous effective charge ($+0.77$) is small compared to 
that of Ti and this feature was first attributed to its more ionic
character~\cite{Zhong94}. This ionicity is inherent to the Harrison
model~\cite{Harrison80} and was confirmed in some ab initio
studies~\cite{Cohen92a,Cohen92b}. Surprisingly, our decomposition reveals
however that the anomalous charges of the O 2s  ($+0.73$) and O 2p 
($+1.50$) bands are not small at all. They are nevertheless roughly
compensated by another Ba 5s  ($+0.11$)and Ba 5p  ($+1.38$) anomalous
contributions. This result suggests that there are dynamical changes of
hybridization between Ba and O orbitals as it was the case between O and Ti,
except that the mechanism is here restricted to {\it occupied} states. {\it This basically
corresponds to a unitary transform within the subspace of the occupied states which is unable
to displace the global Wannier center of the valence charge}. Our result so supports the
hybridization of Ba orbitals, in agreement with experiment~\cite{Nemoshkalenko85,Hudson93},
LCAO calculations~\cite{Pertosa78a,Pertosa78b} and DFT~\cite{Weyrich85a} computations.
Similar compensating contributions were recently  observed in ZnO which has conventional Born
effective charges~\cite{Massidda95} and in a series of alkaline-earth
oxides~\cite{Posternak97}.

We note that a confusion sometimes appears that should be removed: the
{\it amplitude} of the anomalous contributions to $Z^{*(T)}$ is not related
to the amplitude of the hybridizations but to the {\it rate of change} of these
hybridizations under atomic displacements. It is clear that, in BaTiO$_3$, the
Ba 5p contribution to the O 2p bands is smaller than the contribution from
the Ti 3d orbitals~\cite{Weyrich85a,Cohen92a}. However, the high
sensitivity of this relatively weak covalent character under atomic positions is
sufficient to produce large band by band anomalous contributions to $Z^{*(T)}$.
{}From that point of view, 
the Born effective charge  appears therefore as a sensitive tool to identify
the presence of even small hybridizations.

Finally, concerning the oxygen, even if O$_\parallel$ and O$_\perp$ are
defined respectively for a displacement of O  in the Ti and Ba direction, it
seems only qualitative to associate $Z^{*(T)}_{O_\parallel}$ with $Z^{*(T)}_{Ti}$
and $Z^{*(T)}_{O_\perp}$ with $Z^{*(T)}_{Ba}$ as suggested in Ref.~\cite{Zhong94}.
The O 2{p} anomalous contributions to Ti and O$_\parallel$ do not exactly
compensate. Moreover, O 2{p} contribution to $Z^{*(T)}_{Ba}$ does not come
from O$_\perp$ only but has equivalent contributions from O$_\parallel$.
This seems to confirm the idea of Bennetto and Vanderbilt~\cite{Bennetto96}
that in 3D materials, transfers of charges are not necessarily restricted
to a particular bond, but is a rather complex mechanism that must be treated
as a whole.

To summarize, our study has clarified the mixed ionic-covalent character of
BaTiO$_3$: it clearly establishes that the covalent character is not
restricted to the Ti-O bond but also partly concerns the Ba atom.  Moreover,
it leads to a more general issue. It illustrates that {\it the presence
of a large anomalous charge requires a modification of the interactions
between occupied and unoccupied electronic state. The contributions
originating from the change of the interactions between two occupied
states  correspond to unitary transforms within the subspace of the 
valence charge~: they compensate, and do not modify the global value of $Z^{*(T)}$}.

     \subsection{SrTiO$_3$}

\begin{figure}[tb]
\caption{Kohn-Sham electronic band structure of SrTiO$_3$.}
\label{Fig.b.Sr}
\vspace{2cm}
\centerline{\psfig{figure={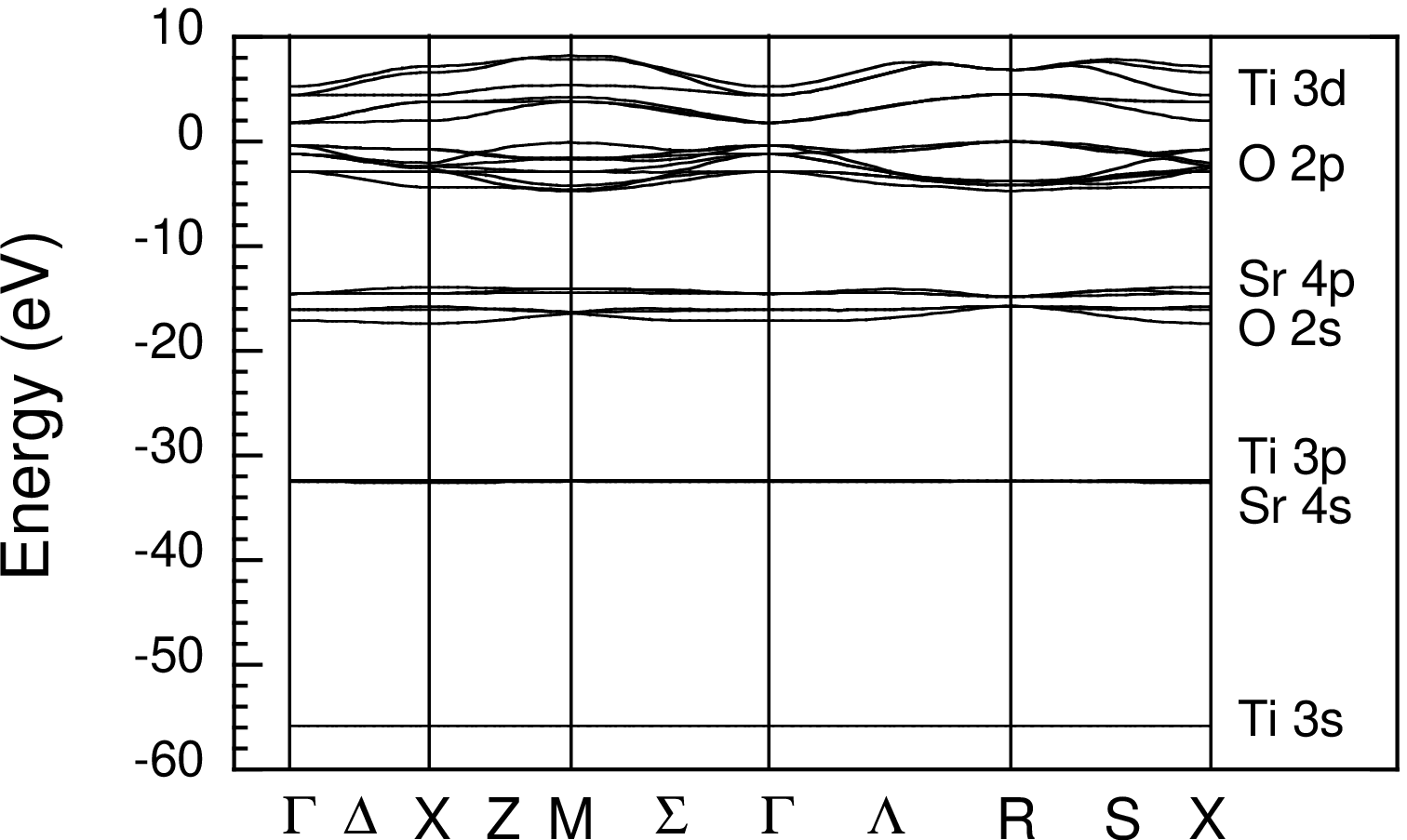},angle=0,height=6.0cm}}
\vspace{4cm}
\end{figure}

The same analysis is now performed on SrTiO$_3$.  Its band structure (Fig.
\ref{Fig.b.Sr}) 
is very similar to that of BaTiO$_3$, except that the Ti 3p  and
Sr 4s bands are energetically very close to each others. Consequently, they
strongly mix and it should be relatively meaningless to separate their
respective contributions. The Sr 4p and O 2s states are in the same
energy region but can be separated, contrary to what was observed in a
study of SrO~\cite{Posternak97}.

The result of the decomposition is very similar (Table \ref{Z*bbbSr})
to that reported for BaTiO$_3$. There is still a giant contribution to
$Z^{*(T)}_{Ti}$ from the O 2p  bands. On the other hand, while the Ba 5p  bands
were approximately centered between O 2s  and O 2p  bands in BaTiO$_3$,
the Sr 4p  electrons are closer to the O 2s  bands and mainly hybridize with
them in SrTiO$_3$. This phenomenon produces large but compensating
contributions from Sr 4p  and O 2s  bands to $Z^{*(T)}_{Sr}$. Such an evolution
is in agreement with the picture that anomalous contributions originate from off-site
orbital hybridization changes.

     \subsection{Other examples}

{}From the two previous results that concern two very similar materials, it might be suggested
that not only the dynamical hybridization of the valence bands with unoccupied d-states but also
the particular cubic perovskite structure of ABO$_3$ compounds plays a major role in determining
$Z^{*(T)}$.  For instance, the anomalous charge could  partly originate in
the local fields at the atomic sites, known to be anomalously large in this
structure~\cite{Slater50}

It is interesting to observe that anomalous charges are not restricted to perovskite
solids but were also detected in a series of alkaline-earth oxides  of rocksalt structure
(CaO, SrO, BaO)~\cite{Posternak96,Posternak97} or even
Al$_2$Ru~\cite{Ogut96,Rabe96}, all examples where the unoccupied d-states
seem to play a major role. Interestingly, two materials belonging to the same
structure can present completely different charges. This was illustrated for 
the case of TiO$_2$ rutile and SiO$_2$ stishovite~\cite{Lee94a,Lee94b}: 
while relatively conventional charges were observed on Si
(+4.15) and O (-2.46) along the Si-O bond in stishovite, giant effective charges,
similar to those of BaTiO$_3$, were obtained on Ti (+7.33) and O (-4.98) along
the Ti-O bond in rutile. Similarly, no anomalous charge was reported for MgO
($Z^{*(T)}_O=-2.07$), presenting the same rocksalt structure than BaO
($Z^{*(T)}_O=-2.80$)~\cite{Posternak97}. In the same spirit, the same atom 
in different environments can present similar dynamical charge, as illustrated 
for $Z^{*(T)}_{Ti}$ in BaTiO$_3$ and TiO$_2$~\cite{Lee94b}, or for $Z^{*(T)}_{Zr}$ 
in BaZrO$_3$~\cite{Zhong94} and ZrO$_2$~\cite{DetrauxU}. Also, in the family 
of ABO$_3$ compounds, giant effective charges are observed on Ti in CaTiO$_3$ 
($Z^{*(T)}_{Ti}=7.08$,~\cite{Zhong94})  but not on Si in CaSiO$_3$ ($Z^{*(T)}_{Si}=4.00$,
\cite{Stixrude97}).

We observe that the presence of partly hybridized d-states
seems the only common feature between the materials presenting giant
anomalous effective charges, listed up to date. This feature finds  a
basic justification  within the BOM of Harrison: the interaction parameters
involving d-states are indeed much more sensitive to the interatomic
distance than those involving, for example, s and p orbitals~\cite{Harrison80}:
They will therefore be associated to larger dynamical transfer of charge and
will generate higher $Z^{*(T)}$.

\section {Sensitivity of $Z^{*(T)}$ to structural features}
\label{Sensitivity}

In the litterature, calculations of $Z^{*(T)}$ essentially focused on the cubic phase
of ABO$_3$ compounds
\cite{Resta93,Ghosez94,Zhong94,Ghosez95a,Posternak94,Ghosez95b,Rabe94,Yu95}.
On the basis of an  early study of KNbO$_3$ \cite{Resta93}, it was concluded
that the Born effective charges are independent of the ionic ferroelectric
displacements (i.e. they remain similar in the different phases).  Another
investigation in the tetragonal phase of KNbO$_3$ and PbTiO$_3$
\cite{Zhong94}, seemed to confirm that $Z^{*(T)}$ are quite insensitive to
structural details. 

These results were surprising if we remember that anomalous contributions to $Z^{*(T)}$ are closely
related to orbital hybridizations, these in turn, well known to be strongly affected by the phase
transitions~\cite{Cohen92a,Cohen92b}. We will see in this Section that, contrary to what was
first expected, Born effective charges in BaTiO$_3$ are strongly dependent of the structural
features. 

We first investigate the sensitivity of the Born effective charges to the
ferroelectric atomic displacements~\cite{Ghosez95b}. For that purpose, we
compute $Z^{*(T)}$ in the three ferroelectric phases at the experimental unit cell
parameters, with relaxed atomic positions as reported in Ref.~\cite{GhosezU}. 
Table \ref{Z*ferro-BTO} 
summarizes the results for a cartesian set of axis where the $z$-axis points
in the ferroelectric direction. The Ba and Ti charge tensors are diagonal in each
phase for this particular choice. In the case of O, we note the presence of a
small asymmetric contribution for the lowest symmetry phases. The
eigenvalues of the symmetric part of the tensor are also reported. In each
phase, the eigenvector associated to the highest eigenvalue of O approximately
points in the Ti-O direction and allows to identify the highest contribution as
O$_{\parallel}$. The other eigenvalues can be referred to as O$_{\perp}$, by
analogy with the cubic phase. 

Although the charges of Ba and O$_{\perp}$ remain globally unchanged in the 4
phases, strong modifications are observed for Ti and O$_{\parallel}$:
for example, changing the Ti position by 0.076\AA \, (2\% of the unit cell
length) when going  from  the cubic to the rhombohedral phase, reduces the
{\it anomalous} part of $Z^{*(T)}_{Ti}$ by more than 50\% along the ferroelectric
axis (Table~\ref{Z*ferro-BTO}). Equivalent evolutions are observed in the
other ferroelectric phases. Similar changes were detected 
in KNbO$_3$~\cite{Wang96b}. 

\begin{figure}[tb]
\caption{Born effective charge of Ti atoms in the direction of the shortest Ti-O
bond length ($d_{min}$) as a function of this interatomic distance, for the cubic
(square), tetragonal (lozenge), orthorhombic (circle) and rhombohedral
(triangle) phases.}
\label{Fig.Z*TiO}
\vspace{2cm}
\centerline{\psfig{figure={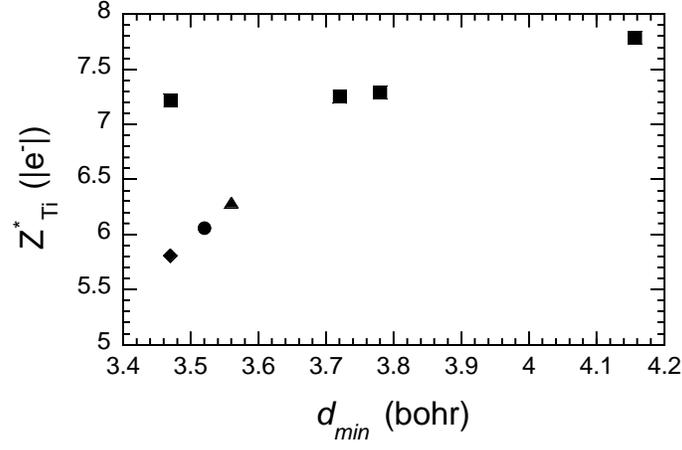},angle=0,height=6.0cm}}
\vspace{4cm}
\end{figure}

In the isotropic cubic structure, Harrison had explained the large value of
$Z^{*(T)}$ in terms of the Ti-O bond length. For the anisotropic ferroelectric
phases, it should be intuitively expected that the shortest Ti-O distance
$d_{min}$ in the structure will dominate the bonding properties. It is
therefore tempting to transpose the Harrison model to understand the
evolution of $Z^{*(T)}$ in terms of the distance $d_{min}$. 
In Fig.~\ref{Fig.Z*TiO}, the amplitude of $Z^{*(T)}_{Ti}$ in 
the direction of the shortest  Ti-O bond length of each phase is
plotted with respect to  $d_{min}$. I similar graph can be obtained for O.
For the different phases, at the experimental lattice parameters, we observe 
that the anomalous parts evolve quasi linearly with $d_{min}$.

Independently from the previous calculations, we also investigated the
evolution of $Z^{*(T)}$ under isotropic pressure (Table \ref{Z*pressure}).  
In contrast with the changes observed with respect to the atomic
displacements, the charge appears essentially insensitive to isotropic
compression. In particular, in the compressed cubic cell at 3.67 \AA\ where
the Ti-O  distance is {\it the same} that the shortest Ti-O bond length  in the
tetragonal structure~\cite{Note_T}, $Z^{*(T)}_{Ti}$  remains very close to its 
value at the optimized volume. This new element clearly invalidates the expected
dependence from $Z^{*(T)}$ to $d_{min}$. 

The fundamental difference between the cubic and tetragonal structures
lies in the fact that in the cubic phase every Ti-O distance is equal to the
others, while in the tetragonal phase, along the ferroelectric axis, a short
Ti-O bond length ($d_{min}$) is followed by a larger one ($d_{max}$) which
{\it breaks} the Ti-O chain in this direction.  In order to verify that it is not
this large Ti-O distance which, alternatively to $d_{min}$, is  sufficient to
inhibit the giant current associated to the anomalous charges, we also
performed a calculation in an expanded cubic phase where
$a_o= 2.d_{max}$: we observe however that the Ti charge is even larger than
in the optimized cubic phase.

We conclude from the previous investigations that {\it the amplitude of $Z^{*(T)}$
in BaTiO$_3$ is not dependent on a particular interatomic distance
($d_{min}$, $d_{max}$) but is more critically affected by the 
anisotropy of the Ti environment along the Ti--O chains}. In
agreement with this picture, Wang {\it et al.}~\cite{Wang96b} reported
recently an insensitivity of $Z^{*(T)}$ to a tetragonal macroscopic strain in
KNbO$_3$. Also, the charges reported by Bellaiche {\it et al.}~\cite{BellaicheU} 
in mixed a compound as PZT, where the ionic environment becomes anisotropic,
seem to confirm our results.

A band by band decomposition of $Z^{*(T)}_{Ti}$ (Table~\ref{Z*Ti})  
points out that the difference between the
cubic and tetragonal phases is essentially localized at the level of the O 2p
bands (+1.48 instead of +2.86) while the other contributions remain very
similar.  This suggests an intuitive explanation. In the cubic phase the O 2p 
electrons are widely delocalized and dynamical transfer of charge can
propagate along the Ti-O chain as suggested by Harrison. In the tetragonal
phase, the Ti-O chain behaves as a sequence of Ti-O dimers for which the
electrons are less polarizable. This smaller polarizability is confirmed by 
a similar reduction of the optical dielectric constant along the ferroelectric direction.
This analysis seems plausible from the Wannier function 
analysis reported recently by Marzari and Vanderbilt~\cite{MarzariU}.

Finally, let us mention that if the evolution of $Z^{*(T)}$  is relatively weak
under isotropic pressure, it would be wrong to consider that the dynamical
properties of BaTiO$_3$ are insensitive to the volume: small changes are
observed that are of the same order of magnitude than for other compounds
like SiC~\cite{Wang96a,Wellenhofer96}. The direction of the evolution
is however different. Moreover, the evolution of the different charges is even not
identical: while the absolute value of $Z^{*(T)}_{Ba}$ and $Z^{*(T)}_{O \perp}$ decreases
with increasing volume,  the inverse behaviour is observed for $Z^{*(T)}_{Ti}$ and
$Z^{*(T)}_{O\parallel}$. 

Here also, the band by band decomposition (Table~\ref{Z*Ba}) 
reveals some hidden features. In the compressed cubic phase, the anomalous
part of the Ba 5p , Ba 5s  and Ti 3p  bands are 50\% larger than in the
optimized cubic cell. This suggests an evolution of the interactions between
occupied orbitals that is coherent with the modification of the interatomic
short-range forces observed independently~\cite{Ghosez96}. At the opposite,
in our expanded cubic phase, most of the anomalous contributions to
$Z^{*(T)}_{Ba}$ and $Z^{*(T)}_{Ti}$ have disappeared in agreement with the picture of a
more ionic material. The O 2p  contribution, is the only one that remains
surprisingly large. Comparing to the value obtained for the cubic phase at 
the experimental volume, its evolution was even more important than the linear
dependence upon the bond length, expected from the Harrison model.

\section {Spontaneous polarization}
\label{Spontaneous}

The spontaneous polarization ($P_s$) of the ferroelectric phases can be
determined by integrating the change of polarization along the path of atomic
displacement from the paraelectric cubic phase (taken as reference) to the
considered ferroelectric structure. If the effective charges were roughly
constant, this integration should be approximated by:
\begin{equation}
\label{Eq.Ps}
P_{s, \alpha} = \frac{1}{\Omega_o} \; \sum_{\kappa, \beta} \;
Z^{*(T)}_{\kappa, \alpha \beta} \; \delta \tau_{\kappa,\beta}
\end{equation}
However, we have seen, in the previous Section, that the Born effective
charges are strongly affected by the atomic displacements. It is therefore
important to investigate their evolution all along the path of atomic
displacements from one structure to the other.

\begin{figure}[tb]
\caption{Evolution of the amplitude of $Z^{*(T)}_{Ti}$ in the $<111>$ direction all
along the path of atomic displacements from the cubic ($\lambda=0$) to the rhombohedral
($\lambda=1$) phase. The distortion of the cubic cell has been neglected.}
\label{Fig.Z*.cr}
\vspace{2cm}
\centerline{\psfig{figure={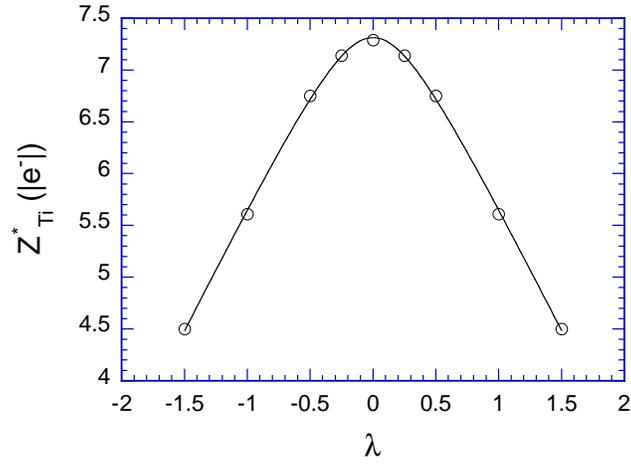},angle=0,height=6.0cm}}
\vspace{4cm}
\end{figure}

We performed the calculation for a transformation from the cubic to the
rhombohedral structure. The rhombohedral macroscopic strain is very small
and was neglected~\cite{NotePs}~: our calculation was performed  by displacing the atoms
to their theoretically optimized position in rhombohedral symmetry, 
when keeping the cubic lattice parameters. The result is reported in 
Figure~\ref{Fig.Z*.cr}, for $Z^{*(T)}_{Ti}$ along the ferroelectric direction. 
A similar curve can be
obtained for $Z^{*(T)}_{O \parallel}$. We observe that the evolution of $Z^{*(T)}$ is
approximately quadratic close to the cubic phase. However, it becomes
rapidly linear, and remains linear for displacements even larger than those
associated to the ferroelectric distortion.   

Expecting a similar evolution of the dynamical charges for the tetragonal and
orthorhombic displacements, an estimation of the spontaneous polarization
in the ferroelectric phases can be found when using Eq.~(\ref{Eq.Ps}) with a
mean effective charge determined from its value in both phases. Using a mean charge
estimated from the values in the para- and ferro-electric phases, we obtain the 
spontaneous polarizations presented in Table \ref{Ps}. 

Our results are only in relative agreement with the
experiment~\cite{Wieder55,Hewat73} and suggest different comments.
Firstly, we would like to mention that part of the discrepancy must be
assigned to the theoretical overestimation of the computed ferroelectric displacements,
discussed in Ref.~\cite{GhosezU}~: when using the experimental
displacements of Ref.~\cite{Hewat74}, we recover a better estimation of
$P_s$ as in Ref.~\cite{Zhong94}. The dispersion of X-rays diffraction data
makes however difficult the exact identification of the ferroelectric
displacements. Secondly, another part of the error could be due to the lack of
polarization dependence of the LDA~\cite{Gonze95}. Finally, we note that there 
is also some uncertainty on the experimental value of $P_s$. 

\section {Conclusions}
\label{Ccls}

In this paper, we first analyzed the links between different definitions of atomic charge.
We have shown that, contrary to the static definitions, dynamical effective charges
also depend on the electronic charge reorganisation induced by an atomic displacement. 
The amplitude of this dynamical contribution is monitored not only by the bonding with the
other atoms but also, for large systems, by the condition imposed on
the macroscopic electric field. A unified treatment of the concept of dynamical charge
in molecules, large clusters, and truly periodic systems has been presented, in which
the Born effective charge and the optical dielectric constant appear as the two fundamental
quantities. The microscopic origin of the dynamical contribution has been discussed
in terms of local polarizability and delocalized transfers of electrons. 

Based on various first-principles results, we have then emphasized that the Born effective
charges are anomalously large in the family of ABO$_3$ compounds: their
amplitude can reach more than twice that of the nominal ionic charges. This
feature was explained in terms of interatomic transfers of charge, produced
by ``off-site'' dynamical changes of hybridization. For BaTiO$_3$ and SrTiO$_3$, we have
brought to light complex dynamical changes of hybridization,  concerning not
only Ti and O but also Ba and Sr orbitals. The hybridizations restricted to
occupied states generate however compensating anomalous contributions so
that, finally, the total value of $Z^{*(T)}$ is essentially affected by dynamical
changes of hybridization between O 2p and Ti 3d orbitals.

As a more general issue, it appears that the existence of partial
hybridizations between occupied and unoccupied states is an important
feature for candidate to large anomalous Born effective charges. Moreover,
the dynamical transfers of charge are expected to be larger when such a
hybridization involve d states, for which the interactions parameters with
other orbitals are particularly sensitive to the interatomic distance.

Investigating the evolution of $Z^{*(T)}$ to the structural features, we have
shown that they are strongly affected by the ferroelectric atomic
displacements and much less sensitive to isotropic pressure. The
results have clarified that the amplitude of $Z^{*(T)}$ is not monitored by a
particular interatomic distance but is dependent on the anisotropy of the Ti
environment along the Ti-O chains.

Finally, the effective charges were used to estimate the spontaneous
polarization in the ferroelectric phases of BaTiO$_3$. For that purpose, their
evolution was investigated all along the path of atomic displacement from
the cubic to the rhombohedral structure.

All along this work, we only focused on the {\it microscopic} mechanisms
that govern the amplitude of the Born effective charges.
In independent studies, it was also emphasized that the anomalously large Born
effective charges produce a giant LO-TO splitting in  ABO$_3$ compounds,
specially for the ferroelectric phonon mode~\cite{Zhong94,Ghosez96}. 
Moreover, it was demonstrated that this feature is associated to the existence of an
anomalously large destabilizing dipole-dipole interaction, sufficient to
compensate the stabilizing short-range forces and induce the ferroelectric
instability~\cite{Ghosez96}. In materials where polar modes play a major role, 
the Born effective charge appears therefore also as a ``key concept'' to relate the 
electronic and structural properties~\cite{GhosezC}. 

\acknowledgments
Ph. G. would like to thank 
R. Resta for numerous illuminating discussions, 
Ph. Lambin for his help in shell-model calculations, 
L. L. Boyer for clarifying some aspects of the SCAD model, 
and N. Marzari and D. Vanderbilt for providing an early copy of their recent results. 
X.G. is grateful to the National Fund for Scientific Research (FNRS-Belgium)  for financial support. 
We acknowledge the use of IBM-RS 6000 work stations from common projects between 
IBM Belgium, UCL-PCPM  and FUNDP-SCF, as well as the use of the 
Maui High Performance Computing Center IBM-SP2.  
We thank Corning Incorporated for the availability of the PlaneWave code,  as well as 
J.-M. Beuken for his kind and permanent computer assistance. 
We also acknowledge financial support from the PAI-UIAP P4/10. 

\vspace{2mm}

\appendix

\section{The Harrison model}
\label{Apdx-2}

In this Appendix, we briefly describe the bond-orbital model proposed by
Harrison for the case of ABO$_3$ compounds~\cite{Harrison80}.  
In particular, we pay a particular
attention to the definition and the calculation of static and dynamical
charges within this model. Values are reported for SrTiO$_3$.

The bond orbital model of Harrison consists in a simplified tight-binding
model, where the Hamiltonian is limited to the on-site and
nearest-neighbour terms. Moreover, for ABO$_3$ compounds, it is assumed
that the A atom has no other function than to provide electrons to the
system, and is fully ionized. The only considered interactions involve 
B and O atom orbitals.

The model includes O 2s, O  2p  and B  d  orbitals, interacting through
$V_{sd,\sigma}$, $V_{pd,\sigma}$ and $V_{pd,\pi}$ parameters. As the matrix
elements $V_{sd,\sigma}$ and $V_{pd,\sigma}$ are nearly
identical, it is suggested to construct two $sp$ hybrids on
the oxygen, $|h_{\pm}>=(|s> \pm |p>)/ \sqrt 2$.  Each of these hybrid will have a large matrix
element ($V^+_{hd,\sigma}=  [V_{sd,\sigma}+V_{pd,\sigma}]/ \sqrt 2$) coupling it to the $d$ state
on one side and a negligible matrix element  ($V^-_{hd,\sigma}=  [V_{sd,\sigma}-V_{pd,\sigma}]/
\sqrt 2$) coupling it to the $d$ state on the other. 

     \subsection{Effective static charge}

In absence of orbital interactions, the static charges of SrTiO$_3$ would be of $+2$ on Sr,
$+4$ on Ti and $-2$ on O. However, due to the Ti-O orbital interactions, the transfer of
electrons from Ti to O is not complete. Within the BOM, the effective static charge on O
can be estimated as:
\begin{equation}
Z_O = -2 + T_{\sigma} + T_{\pi}
\end{equation}
where 
\begin{eqnarray}
T_{\sigma}= 4  \Biggl( \frac{V^+_{hd,\sigma}}{[\epsilon_d - 
                                                          (\epsilon_s + \epsilon_p)/2]} \Biggr)^2
\end{eqnarray}
(``4'' because there are 2 hybrids composed of 2 electrons that each interacts
mainly with one Ti neighbour) and
\begin{eqnarray}
T_{\pi}= 8 \Biggl( \frac{V^+_{pd,\pi}}{[\epsilon_d - \epsilon_p)]} \Biggr)^2
\end{eqnarray}
(``8'' because there are 4 electrons that are each partly delocalized on the 2 Ti
neighbours).

For SrTiO$_3$, from the parameters of Matheiss (Ref.~\cite{Harrison80}, p. 445), 
$T_{\sigma} = 0.35$ and
$T_{\pi}= 0.68$ so that the effective static charge on the oxygen atom is $Z_O = 0.96$.

     \subsection{Born effective charge}

Following the discussion of Section III-B, the Born effective charge can be obtained by adding to
the static charge the dynamical contribution  originating in the dependence of the
matrix elements $V$ upon the interatomic bond length $d$.  {}From Harrison's Solid State Table
(see also Eq. (19-11) in Ref.~\cite{Harrison80}), the previous
matrix elements have all the same interatomic dependence:
\begin{eqnarray}
V = K d^{-7/2} \\
\delta V = - \frac{7}{2} V \frac{\delta d}{d}
\end{eqnarray}
so that 
\begin{eqnarray}
(V+ \delta V)^2 &=& V^2 + 2 \; V \; \delta V + {\cal O}(\delta V^2) \\
                         &=& V^2 + 2 \; (\frac{-7}{2}) V^2 \frac{\delta d}{d} + {\cal O}(\delta V^2)
\end{eqnarray}

When displacing the O atom along the Ti-O direction, there will be an additional
transfer of electron from O to the nearest Ti that is equal to 
\begin{eqnarray}
\delta T_{\sigma} &=& 2 [ 2 \; (\frac{-7}{2}) (\frac{V_{hd,\sigma}}{\epsilon_d -
                                          \epsilon_h})^2 ] \frac{\delta d}{d}
\\
\delta T_{\pi} &=&   4 [ 2 \; (\frac{-7}{2}) (\frac{V_{pd,\pi}}{\epsilon_d -
                                          \epsilon_p})^2 ] \frac{\delta d}{d}
\end{eqnarray}
On the other hand, there will be the same transfer of charge from the other
neighbour Ti atom to O, so that the previous electrons are globally transferred
on a distance
$2d$. The change of polarization associated to this transfer of charge is:
\begin{eqnarray}
\delta P_{\sigma} &=& - 28 (\frac{V_{hd,\sigma}}{\epsilon_d -
                                          \epsilon_h})^2  \delta d
\\
\delta P_{\pi} &=&   - 56 (\frac{V_{pd,\pi}}{\epsilon_d -
                                          \epsilon_p})^2 \delta d
\end{eqnarray}
The associated dynamic contribution to the Born effective charge is ($\delta
P/\delta d$):
\begin{eqnarray}
\delta Z^{*(T)}_{\sigma} &=& - 28 (\frac{V_{hd,\sigma}}{\epsilon_d -
                                          \epsilon_h})^2  
\\
\delta Z^{*(T)}_{\pi} &=&  - 56 (\frac{V_{pd,\pi}}{\epsilon_d -
                                          \epsilon_p})^2 
\end{eqnarray}

The effective charge on the O atom for a displacement along the Ti-O direction is
therefore:
\begin{eqnarray}
Z^{*(T)}_{O \parallel} = Z_O + \delta Z^{*(T)}_{\sigma} + \delta Z^{*(T)}_{\pi} 
\end{eqnarray}

For SrTiO$_3$, from the parameters of Matheiss (Ref.~\cite{Harrison80}, p. 445), 
$\delta Z^{*(T)}_{\sigma} = -2.45$ and
$\delta Z^{*(T)}_{\pi}  = -4.76 $ so that $Z^{*(T)}_{O \parallel} = -8.18 $.



\newpage
\newpage
\newpage

\newpage
\begin{table}
\caption {Static charges of BaTiO$_3$ in the cubic structure. }
\label{Zcubic}
\begin{center}
\begin {tabular}{lcccc}
&$Z_{Ba}$ &$Z_{Ti}$ &$Z_{0}$  &Reference \\
\hline
Nominal &$+2$	     &$+4$	     &$-2$	  \\
{Empirical models} 
&$+2.00$  &$+0.19$ &$-0.73$  &Ref. \cite{Harrison80} \\
&$+1.40$  &$+2.20$  &$-1.20$   &Ref.  \cite{Hewat73}    \\
&$+2.00$   &$+1.88$  &$-1.29$      &Ref.  \cite{Michel80} \\
&$+1.86$  &$+3.18$   &$-1.68$   &Ref. \cite{Khatib89} \\
&$+1.48$  &$+1.86$   &$-1.11$   &Ref. \cite{Turik88} \\
&$+2.00$  &$+1.86$   &$-1.29$    &Ref. \cite{Turik95}  \\
{First-principles calculations} 
&$+2.00$   &$+2.89$  &$-1.63$  &Ref. \cite{Cohen90} \\
&$+2.12$   &$+2.43$   &$-1.52$ &Ref. \cite{Xu90}   \\
&$+1.39$  &$+2.79$   &$-1.39$    &Ref. \cite{Xu94} \\
\end{tabular}
\end{center}
\end{table}
\begin{table}
\caption {Born effective charges of BaTiO$_3$ in the cubic structure. }
\label{Z*cubicBa}
\begin{center}
\begin {tabular}{lcccccc}
& &$Z^{*(T)}_{Ba}$  &$Z^{*(T)}_{Ti}$    &$Z^{*(T)}_{O_{\perp}}$   
&$Z^{*(T)}_{O_{\parallel}}$  &Reference \\
\hline
Nominal        &        &$+2$      &$+4$      &$-2$      &$-2$     \\
Experiment     &      &$+2.9$   &$+6.7$   &$-2.4$    &$-4.8$    &Ref.~\cite{Axe67} \\
Models 
&(Shell model)          &$+1.63$   &$+7.51$   &$-2.71$    &$-3.72$   &Ref. \cite{Ghosez95b} \\
&(SCAD model )          &$+2.9$    &$+7.3$    &$-2.2$        &$-5.8$     &Ref.  \cite{Boyer96}   \\
First-principles 
&(Linear response)     &$+2.77$   &$+7.25$   &$-2.15$    &$-5.71$    &Present  \\
&(Berry phase)           &$+2.75$   &$+7.16$   &$-2.11$    &$-5.69$    &Ref. \cite{Zhong94}
\end{tabular}
\end{center}
\end{table}
\begin{table}
\caption {Born effective charges of various ABO$_3$ compounds in their
cubic structure.}
\label{Z*cubicAB}
\begin{center}
\begin {tabular}{lccccl}
ABO$_3$   &$Z^{*(T)}_{A}$ &$Z^{*(T)}_{B}$  &$Z^{*(T)}_{O \parallel}$  
&$Z^{*(T)}_{O \perp}$  &Reference  \\
\hline
nominal      &2    &4        &   -2      & -2   \\
CaTiO$_3$    &2.58  &7.08   &-5.65  &-2.00 &Ref.~\cite{Zhong94}  \\
SrTiO$_3$  &2.56  &7.26  &-5.73  &-2.15 &Present \\
           &2.54 & 7.12 & -5.66 &-2.00 & Ref.~\cite{Zhong94}  \\
 	   &2.55 & 7.56 & -5.92  &-2.12 & Ref.~\cite{LaSota97} \\
           &2.4   &7.0   &-5.8   &-1.8  & Ref.~\cite{Axe67} \\
BaZrO$_3$  &2.73 &6.03 &-4.74 &-2.01 &Ref.~\cite{Zhong94}\\
PbTiO$_3$  & 3.90 & 7.06 & -5.83 &-2.56 & Ref.~\cite{Zhong94}  \\
PbZrO$_3$  & 3.92 & 5.85 & -4.81 &-2.48 & Ref.~\cite{Zhong94}  \\
\hline
nominal    &1    &5        &   -2      & -2   \\
NaNbO$_3$  & 1.13 & 9.11 & -7.01 &-1.61 & Ref.~\cite{Zhong94}  \\
KNbO$_3$   & 0.82 & 9.13 & -6.58 &-1.68 & Ref.~\cite{Resta93} \\
           &1.14 & 9.23 & -7.01 &-1.68  & Ref.~\cite{Zhong94}  \\
           & 1.14 & 9.37 & -6.86 &-1.65 & Ref.~\cite{Yu95} \\
\hline
nominal    &-  & 6    &   -2  & -2   \\
WO$_3$     &-  & 12.51 & -9.13 &-1.69    &Ref.~\cite{Detraux97}  
\\
\end{tabular}
\end{center}
\end{table}
\begin{table}
\caption {Band by band decomposition of $Z^{*(T)}$ in the
optimized cubic phase of BaTiO$_3$. The contributions have been separated into
a reference value and an anomalous charge (see text).}
\begin{center}
\label{Z*bbbBa}
\begin {tabular}{lrrrrr}
Band 
  &{$Z_{Ba}^{(T)}$}   
 &{$Z_{Ti}^{(T)}$} 
 &{$Z_{O_{\perp}}^{(T)}$} 
 &{$Z_{O_{\parallel}}^{(T)}$} 
 & Total \\
\hline
$Z_{\kappa}$
 & {$+10.00$}
 &{$+12.00$}
 &{$+6.00$}
 &{$+6.00$}
 & $+40$ \\
Ti 3s   &$0+0.01$   &$-2-0.03$    &$0+0.00$   
 &$0+0.02$   &$-2.00 $\\
Ti 3p    &$0+0.02$   &$-6-0.22$    &$0-0.02$  
  &$0+0.21$   &$-6.03 $\\
Ba 5s    &$-2-0.11$    &$0+0.05$    &$0+0.02$  
  &$0+0.01$   &$-2.01 $\\
O 2s     &$0+0.73$    &$0+0.23$    &$-2-0.23$  
  &$-2-2.51$   &$-6.01 $\\
Ba 5p   & $-6-1.38$   &$0+0.36$    &$0+0.58$  
 &$0-0.13$   &$-5.99 $\\
O 2p    &$0+1.50$   &$0+2.86$   &$-6-0.50$  
  &$-6-3.31$  &$-17.95$\\
\hline
Total        
 &{$+2.77$}
 &{$+7.25$}
 &{$-2.15$}
 &{$-5.71$}
 &$+0.01$ \\
\end{tabular}
\end{center}
\end{table}
\begin{table}
\caption {Band by band decomposition of $Z^{*(T)}$ in the
experimental cubic phase of SrTiO$_3$. The contributions have been separated
into a reference value and an anomalous charge (see text).}
\label{Z*bbbSr}
\begin{center}
\begin {tabular}{lrrrrr}
\begin{tabular}{l}
Band 
\end{tabular}
&{$Z_{Sr}^{(T)}$}   
&{$Z_{Ti}^{(T)}$} 
&{$Z_{O_{\perp}}^{(T)}$} 
&{$Z_{O_{\parallel}}^{(T)}$} 
& Total \\
\hline
\begin{tabular}{l}
$Z_{\kappa}$  
\end{tabular}
&{$+10.00$}
&{$+12.00$}
&{$+6.00$}
&{$+6.00$}
&$+40$ \\
\begin{tabular}{l}
Ti 3s   
\end{tabular}
 &$0+0.01$   &$-2-0.03$  &$0+0.00$   &$0+0.03$ &$-1.99 $\\
\begin{tabular}{ll}
Sr 4s  \\
Ti 3p
\end{tabular}
$\Bigr\}$ 
&$-2+0.02$  &$-6-0.18$  &$0-0.03$ &$0+0.23$ &$-7.99 $\\ 
\begin{tabular}{l}
O 2s   
\end{tabular}
&$0+3.08$   &$0+0.02$  &$-2-1.31$  &$-2-0.48$ &$-6.00 $\\
\begin{tabular}{l}
Sr 4p  
\end{tabular}
&$-6-3.11$   &$0+0.37$  &$0+1.42$   &$0-0.10$ &$-6.00 $\\
\begin{tabular}{l}
O 2p   
\end{tabular}
&$0+0.56$   &$0+3.08$  & $-6-0.12$ &$-6-3.41$ & $-18.01$\\
\hline
\begin{tabular}{l}
Total 
\end{tabular}       
&{$+2.56$}
&{$+7.26$}
&{$-2.15$}
&{$-5.73$}
&$+0.01$ \\
\end{tabular}
\end{center}
\end{table}
\begin{table}
\caption {Born effective charges in the three ferroelectric phases of BaTiO$_3$. 
Tensors are reported in cartesian coordinates, with the $z$-axis along the
ferroelectric direction. For Ba and Ti, the tensors are diagonal and only the
principal elements are mentioned. For O, full tensors are reported. The
eigenvalues of the symmetric part of $Z^{*(T)}$ are mentioned in brackets; the
eigenvector associated to the highest eigenvalue approximately points in the Ti
direction. In the cubic phase, we had: $Z^{*(T)}_{Ti}=7.29$, $Z^{*(T)}_{Ba}=2.74$,
$Z^{*(T)}_{O \parallel}=-5.75$ and $Z^{*(T)}_{O \perp}=-2.13$. }
\label{Z*ferro-BTO}
\begin{center}
\begin {tabular}{lccc}
&Tetragonal &Orthorhombic &Rhombohedral \\
\hline
\\ 
$\begin{array}{l}
{Z^{*(T)}_{Ba}}
\end{array}$
&
$(
\begin{array}{lcr}
{+2.72}&{+2.72}&{+2.83}
\end{array}
)$
&
$(
\begin{array}{lcr}
{+2.72}&{+2.81}&{+2.77}
\end{array}
)$
&
$(
\begin{array}{lcr}
{+2.79}&{+2.79}&{+2.74}
\end{array}
)$
\\
\\ 
$\begin{array}{l}
{Z^{*(T)}_{Ti}}
\end{array}$
&
$\left(
\begin{array}{lcr}
{+6.94}&{+6.94}&{+5.81}
\end{array}
\right)  $
&
$\left(
\begin{array}{lcr}
{+6.80}&{+6.43}&{+5.59}
\end{array}
\right)  $
&
$\left(
\begin{array}{lcr}
{+6.54} &{+6.54} &{+5.61} 
\end{array}
\right)  $
\\
\\ 
$\begin{array}{l}
{Z^{*(T)}_{O1}}\\
{} \\
{}  
\end{array}$
&
$\left(
\begin{array}{lcr}
{-1.99} &{0} &{0} \\
{0} &{-1.99} &{0} \\
{0} &{0} &{-4.73} 
\end{array}
\right)$  
&
$\left(
\begin{array}{lcr}
{-2.04} &{0} &{0} \\
{0} &{-3.63} &{+1.38} \\
{0} &{+1.57} &{-3.17} 
\end{array}
\right)$  
&
$\left(
\begin{array}{lcr}
{-2.54} &{-0.99} &{+0.63} \\
{-0.99} &{-3.68} &{+1.09} \\
{+0.72} &{+1.25} &{-2.78} 
\end{array}
\right)$  
\\
\\ 
&
$[
\begin{array}{lcr}
{-1.99}&{-1.99}&{-4.73}
\end{array}
]$
&
$[
\begin{array}{lcr}
{-1.91}&{-2.04}&{-4.89}
\end{array}
]$
&
$[
\begin{array}{lcr}
{-1.97} &{-1.98} &{-5.05} 
\end{array}
]$
\\
\\ 
$\begin{array}{l}
{Z^{*(T)}_{O2}}\\
{} \\
{}  
\end{array}$
&
$\left(
\begin{array}{lcr}
{-2.14} &{0} &{0} \\
{0} &{-5.53} &{0} \\
{0} &{0} &{-1.95} 
\end{array}
\right)$  
&
$\left(
\begin{array}{lcr}
{-2.04} &{0} &{0} \\
{0} &{-3.63} &{+1.38} \\
{0} &{+1.57} &{-3.17} 
\end{array}
\right)$  
&
$\left(
\begin{array}{lcr}
{-2.54} &{+0.99} &{+0.63} \\
{+0.99} &{-3.68} &{-1.09} \\
{+0.72} &{-1.25} &{-2.78} 
\end{array}
\right)$
\\
\\ 
&
$[
\begin{array}{lcr}
{-1.95}&{-2.14}&{-5.53}
\end{array}
]$
&
$[
\begin{array}{lcr}
{-1.91}&{-2.04}&{-4.89}
\end{array}
]$
&
$[
\begin{array}{lcr}
{-1.97} &{-1.98} &{-5.05} 
\end{array}
]$
\\
\\ 
$\begin{array}{l}
{Z^{*(T)}_{O3}}\\
{} \\
{}  
\end{array}$
&
$\left(
\begin{array}{lcr}
{-5.53} &{0} &{0} \\
{0} &{-2.14} &{0} \\
{0} &{0} &{-1.95} 
\end{array}
\right)$  
&
$\left(
\begin{array}{lcr}
{-5.44} &{0} &{0} \\
{0} &{-1.97} &{0} \\
{0} &{0} &{-2.01} 
\end{array}
\right)$  
&
$\left(
\begin{array}{lcr}
{-4.25} &{0} &{-1.26} \\
{0} &{-1.97} &{0} \\
{-1.44} &{0} &{-2.78} 
\end{array}
\right)  $
\\
\\ 
&
$[
\begin{array}{lcr}
{-1.95}&{-2.14}&{-5.53}
\end{array}
]$
&
$[
\begin{array}{lcr}
{-1.97} &{-2.01} &{-5.44} 
\end{array}
]$
&
$[
\begin{array}{lcr}
{-1.97} &{-1.98} &{-5.05} 
\end{array}
]$ \\
\end{tabular}
\end{center}
\end{table}
\begin{table}
\caption {Evolution of the Born effective charges of BaTiO$_3$
under isotropic pressure in the cubic phase. }
\label{Z*pressure}
\begin{center}
\begin {tabular}{lcccc}
$a_o$ (\AA) &$Z^{*(T)}_{Ba}$ &$Z^{*(T)}_{Ti}$   &$Z^{*(T)}_{O_{\perp}}$  
&$Z^{*(T)}_{O_{\parallel}}$\\
\hline
3.67    &$+2.95$    &$+7.23$   &$-2.28$   &$-5.61$   \\     
3.94    &$+2.77$    &$+7.25$   &$-2.15$   &$-5.71$   \\
4.00    &$+2.74$    &$+7.29$   &$-2.13$   &$-5.75$   \\
4.40    &$+2.60$    &$+7.78$   &$-2.03$   &$-6.31$   \\
\end{tabular}
\end{center}
\end{table}
\begin{table}
\caption {Band by band decomposition of $Z^{*(T)}_{Ti}$ in 
different structure of BaTiO$_3$. The contributions have been separated into
a reference value and an anomalous charge (see text).}
\label{Z*Ti}
\begin{center}
\begin {tabular}{lrrrr}
Band
&$Z^{*(T)}_{Ti}$  
&$Z^{*(T)}_{Ti}$   
&$Z^{*(T)}_{Ti}$  
&$Z^{*(T)}_{Ti}$
\\
&{(cubic - 3.67 \AA)}   
&{(cubic - 3.94 \AA)} 
&{(tetragonal - exp)} 
&{(cubic - 4.40 \AA)} 
\\
\hline
$Z_{\kappa}$   
&{$+12.00$}
&{$+12.00$}
&{$+12.00$}
&{$+12.00$} \\
Ti 3s    &$-2-0.07$   &$-2-0.03$   &$-2-0.05$  &$-2+0.01$ \\
Ti 3p &$-6-0.43$ &$-6-0.22$  &$-6-0.26$  &$-6-0.07$   \\
Ba 5s &$ 0+0.09$ &$ 0+0.05$  &$ 0+0.05$   &$ 0+0.02$   \\
O 2s  &$ 0+0.27$  &$ 0+0.23$  &$ 0+0.25$  &$ 0+0.19$  \\ 
Ba 5p  &$ 0+0.64$  &$ 0+0.36$   &$ 0+0.34$   &$ 0+0.13$    \\ 
O 2p  &$ 0+2.73$  &$ 0+2.86$   &$ 0+1.48$   &$ 0+3.50$     \\
\hline
Total        
&{$+7.23$}
&{$+7.25$}
&{$+5.81$}
&{$+7.78$} \\
\end{tabular}
\end{center}
\end{table}
\begin{table}
\caption {Band by band decomposition of $Z^{*(T)}_{Ba}$ in 
the optimized cubic phase of BaTiO$_3$ and in an expanded cubic structure. The
contributions have been separated into a reference value and an anomalous
charge (see text).}
\label{Z*Ba}
\begin{center}
\begin {tabular}{lrr}
Band &$Z^{*(T)}_{Ba}$  &$Z^{*(T)}_{Ba}$  \\
&(cubic - 3.94 \AA)   &(cubic - 4.40 \AA) \\
\hline
$Z_{\kappa}$   &$+10.00$  &$+10.00$ \\
Ti 3s          &$ 0+0.01$ &$ 0-0.01$    \\
Ti 3p          &$ 0+0.01$ &$ 0+0.01$    \\
Ba 5s          &$-2-0.11$ &$-2+0.00$    \\
O 2s           &$ 0+0.73$ &$ 0+0.37$    \\
Ba 5p          &$-6-1.38$ &$-6-0.44$    \\
O 2p           &$ 0+1.50$ &$ 0+0.66$    \\
\hline
Total          &$+2.77$   &$+2.59$ \\
\end{tabular}
\end{center}
\end{table}
\begin{table}
\caption {Spontaneous polarization in the three ferroelectric phases of
BaTiO$_3$ in $\mu$C/cm$^2$. The results were deduced from 
Eq.~(\protect\ref{Eq.Ps}) when using either
$Z^{*(T)}$ from the cubic phase (Cubic) or a mean charge (Mean) defined as
($Z^{*(T)}_{mean}= 0.68 \times Z^{*(T)}_{cubic}+ 0.32 \times Z^{*(T)}_{ferro}$).  
Results are reported for
the experimental (Exp) and theoretical (Theo) atomic ferroelectric displacements.}
\label{Ps}
\begin{center}
\begin {tabular}{lccccc}
$Z^{*(T)}$ &Positions  &Tetragonal  &Orthorhombic    &Rhombohedral   &Reference  \\
\hline
--       &--     &26.3   &30.7   &33.5   & Exp. \cite{Wieder55} \\
Cubic    &Exp    &30     &26     &44     & Ref. \cite{Zhong94}   \\
Cubic    &Theo    &36.35  &42.78  &43.30  & Present  \\
Mean     &Theo    &34.02  &39.68  &40.17  & Present  \\
Mean     &Exp    &28.64  &36.11  &--     & Present  \\
\end{tabular}
\end{center}
\end{table}


\begin{thebibliography}{100}
\bibitem{Note_M}
We note the unphysical charge on Ba. It  is the result of a negative Mulliken population and
reflects the inadequacy of the Mulliken population analysis which assumes an equal share of
overlap between each pair of atoms.
\bibitem{Note_1}
Nominal and static charges may differ widely due to covalency effects. As the static charge is ill
defined, one usually prefers to define the anomalous contribution in reference to the nominal
charge. The difference between Born effective charge and static charge is sometimes referred to
as the dynamical contribution.
\bibitem{Note_W}
Depending from the structure and from the interactions, the initial position of the
electron Wannier center is not necessarily on an atom. However, in the cubic 
perovskite structure, symmetry imposes to the Wannier centers to be located on an atom 
so that the anomalous  contribution exactly describes the displacement from such a centered 
position.
\bibitem{Note_T}
In the tetragonal phase, shortened and elongated Ti--O bonds alternate along the 
ferroelectric axis. The shortened bond corresponds to an interatomic distance of 
3.67 \AA\ in our optimized tetragonal structure.
\bibitem{NotePs}
The Born effective charges obtained for
the rhombohedral structure when neglecting the strain (i. e. when keeping  a
cubic unit cell) are the following: $Z^{*(T)}_{Ba,11}= +2.79$, $Z^{*(T)}_{Ba,33}=+2.79$,
$Z^{*(T)}_{Ti,11}= +6.54$, $Z^{*(T)}_{Ti,33}= +5.61$, $Z^{*(T)}_{O,\perp}= -1.97$,
$Z^{*(T)}_{O,\parallel}= -5.05$. These values must be compared to those reported
in Table~\ref{Z*ferro-BTO}, where the rhombohedral strain was taken into
accounts. It can be checked that the effect of this strain is negligible.
\end{thebibliography}
\end{document}